\documentclass[12pt]{article}
\usepackage{epsf}
\usepackage{epsfig}

\def\beq{\begin{equation}}
\def\eeq{\end{equation}}

\def\CR{\nonumber\\}

\begin{document}

\begin{titlepage}
\begin{flushright}
 PITHA 04/19\\
 SI-HEP-2004-13\\ 
 hep-ph/0412400\\
 20 December 2004
\end{flushright}
\vskip1.5cm

\begin{center}
\boldmath
{\Large\bf Exclusive radiative and electroweak $b\to d$ and\\[0.3cm] 
$b\to s$ penguin decays at NLO}
\unboldmath 
\vskip 2cm
{\sc M. Beneke$^a$}, \hspace*{0.1cm}{\sc Th.~Feldmann$^b$}
\hspace*{0.1cm}and\hspace*{0.2cm}{\sc D.~Seidel$^a$}
\vskip .5cm

\vspace{0.7cm}
{\sl ${}^a$Institut f\"ur Theoretische Physik E, RWTH Aachen\\
D--52056 Aachen, Germany\\
\vspace{0.3cm}
${}^b$Theoretische Physik 1, Fachbereich Physik,
  Universit\"at Siegen\\ D-57068 Siegen, Germany}\\
\vspace{3\baselineskip}

\vspace*{0.5cm}

\end{center}

\begin{abstract}
\noindent 
We provide Standard Model expectations for the rare radiative 
decays $B\to K^*\gamma$, $B\to \rho\gamma$ and $B\to\omega\gamma$,  
and the electroweak penguin decays $B\to K^*\ell^+\ell^-$ and 
$B\to\rho\,\ell^+\ell^-$ at the next-to-leading order (NLO), extending 
our previous results to $b\to d$ transitions. We consider branching 
fractions, isospin asymmetries and direct CP asymmetries. For 
the electroweak penguin decays, the lepton-invariant mass spectrum 
and forward-backward asymmetry is also included. 
Radiative and electroweak penguin transitions in $b\to d$ are mainly 
interesting in the search for new flavour-changing neutral current 
interactions, but in addition the $B\to\rho\gamma$ decays provide 
constraints on the 
CKM parameters $(\bar\rho,\bar\eta)$. The potential impact of 
these constraints is discussed.
\end{abstract}

\vskip 2.5cm


\vfill

\end{titlepage}


\section{Introduction}

The radiative and electroweak penguin transitions $b\to D\gamma$ and 
$b\to D \ell^+\ell^-$ ($D=d,s$) provide valuable insight into the 
nature of flavour-changing neutral currents. Induced through quantum 
fluctuations in the Standard Model, they may be strongly influenced by new 
heavy particles. This has been studied extensively for inclusive 
$b\to s \gamma$ decays over the past decade. Some years ago the 
QCD factorization approach to exclusive $B$ 
decays~\cite{Beneke:1999br,Beneke:2000ry} was extended to exclusive 
radiative~\cite{Beneke:2001at,Bosch:2001gv,Ali:2001ez} and electroweak penguin 
decays~\cite{Beneke:2001at}, thus opening the possibility to 
perform detailed studies of isospin breaking and CP asymmetries 
in these decays, as well as to obtain more precise predictions for
branching fractions and forward-backward asymmetries. 

At the time when these papers were written, only the exclusive 
$B\to K^*\gamma$ decays had been observed, and a next-to-leading order
(NLO) analysis could be done only for $B\to K^*\gamma,\,\rho\gamma$ and 
$B\to K^*\ell^+\ell^-$, the case of $B\to \rho\,\ell^+\ell^-$ being 
excluded by the absence of the NLO (2-loop) virtual correction 
to the $b\to d\ell^+\ell^-$ transition. This missing piece of input
has recently been computed \cite{Seidel:2004jh, Asatrian:2003vq}, 
while the formal justification of the QCD factorization method 
in radiative decays has been 
pursued in \cite{Chay:2003kb,Descotes-Genon:2004hd}.
Moreover, the $B$ factories have now put limits on the
$B\to\rho\gamma$ and $B\to\omega\gamma$ branching 
fractions~\cite{Aubert:2004fq,Abe:2004kq}, and 
have performed first measurements of 
$B\to K^*\ell^+\ell^-$~\cite{Aubert:2003cm,Ishikawa:2003cp}
including the lepton invariant mass spectrum and forward-backward
asymmetry~\cite{Abe:2004ir}. These 
theoretical and experimental advances motivate 
the following study of the observables of interest with a complete 
next-to-leading order calculation (except for weak 
annihilation).  

The paper is organized as follows: In Section~\ref{sec:theoryinput} 
we give a brief introduction to the theoretical formalism.
We specify the 
input parameters to the analysis and give the various contributions to
the decay amplitudes in numerical form. The theoretical expectations 
for the relevant observables are summarized in 
Section~\ref{sec:pheno}. For both, $B\to V\gamma$ and $B\to
V\ell^+\ell^-$ decays ($V=K^*,\rho,\omega$), 
we discuss branching fractions (lepton
invariant mass spectrum and partially integrated branching fractions
for $B\to V\ell^+\ell^-$), isospin asymmetries (difference between 
charged and neutral $B$ meson decay) and direct CP asymmetries. 
For  $B\to V\ell^+\ell^-$ the forward-backward asymmetry is 
also included. In Section~\ref{sec:3.5} we collect the constraints 
on the CKM unitarity triangle that can be obtained from 
the branching fractions, isospin  and CP asymmetries in  
$B\to\rho\gamma$ decays. 
The analysis of the radiative decays overlaps with recent work 
of Ali et al.~\cite{Ali:2004hn} and Bosch and 
Buchalla~\cite{Bosch:2004nd}, and we compare our results to theirs 
in the appropriate places. 
We conclude in Section~\ref{sec:conclude}. The 
technical Appendix~\ref{app:a} contains the new decay amplitudes 
for $B\to\rho\,\ell^+\ell^-$ related to the up-quark sector of
the effective weak Hamiltonian. 

\section{Theoretical input}
\label{sec:theoryinput}

\subsection{Formalism}
\label{sec:formalism}

The formalism is described in some detail for $b\to s$ transitions 
in~\cite{Beneke:2001at}, and the extension to $b\to d$ is
straightforward.  In the Standard Model the effective Hamiltonian for 
$b\to d$ transitions can be written as
\begin{equation}
H_{\rm eff} = -\frac{G_F}{\sqrt{2}}\left(
   \lambda_t^{(d)} H_{\rm eff}^{(t)} 
    +\lambda_u^{(d)} H_{\rm eff}^{(u)} \right)
\, + \mbox{h.c.},
\end{equation}
with $\lambda_q^{(d)}=V_{qd}^* V_{qb}$, and  
\begin{eqnarray}
 H_{\rm eff}^{(t)} &=&
C_1\, {\cal O}_1^c + C_2\, {\cal O}_2^c
+\sum_{i=3}^{10} C_i\,{\cal O}_i ,
\nonumber  \\[0.2em]
H_{\rm eff}^{(u)} & = &
 C_1\, ({\cal O}_1^c-{\cal O}_1^u) 
      + C_2\, ({\cal O}_2^c-{\cal O}_2^u).
\end{eqnarray}
For $b\to s$ transitions
$\lambda_{t,u}^{(d)}$ is replaced by $\lambda_{t,u}^{(s)}$, so that the 
term $\lambda_u^{(s)} H_{\rm eff}^{(u)}$ is CKM-suppressed and can
be neglected in practice. The main technical point of the present
paper is to add the decay amplitudes from $H_{\rm eff}^{(u)}$, which 
are relevant to the $b\to d$ case. We have written the effective
Hamiltonian in a form such that $H_{\rm eff}^{(u)}$ involves 
the differences of ``tree'' operators $(\bar c b)(\bar D c)$ and 
$(\bar u b)(\bar D u)$ for $b\to D$ ($D=d,s$) transitions.
We use the operator basis as given in~\cite{Beneke:2001at}.
 The numerical values for the Wilson 
coefficients at $\mu=m_b$ at leading-logarithmic
(LL) and NLL order are collected in Table~\ref{tabWilson}. 
The next-to-next-to-leading logarithmic (NNLL) results for 
$C_{9,10}$ were obtained from the solution to the 
renormalization group equations 
given in~\cite{Beneke:2001at}, including the recent computations 
of the 3-loop mixing of the four-quark
operators into $C_9$ \cite{Gambino:2003zm} and among themselves 
\cite{Gorbahn:2004my}.  The coefficient $C_9$ is now complete at 
NNLL as formally required by our analysis. 

\begin{table}[t]
\centerline{\parbox{14cm}{\caption{\label{tabWilson}
Wilson coefficients at the scale $\mu=4.6\,$GeV in leading-logarithmic
(LL) and next-to-leading-logarithmic order (NLL). Input parameters are 
$\Lambda^{(5)}_{\overline{\rm MS}}=0.220$\,GeV, $\hat m_t(\hat m_t)=170$\,GeV, 
$M_W=80.4$\,GeV and $\sin^2\!\theta_W=0.23$. 3-loop running is used 
for $\alpha_s$. 
}}}
\vspace{0.1cm}
\begin{center}
\begin{tabular}{|l|c|c|c|c|c|c|}
\hline\hline
\rule[-2mm]{0mm}{7mm}
 & ${C}_1$ & ${C}_2$ & ${C}_3$ & ${C}_4$ & ${C}_5$
 & ${C}_6$ \\
\hline
\rule[-0mm]{0mm}{4mm}
LL  & $-0.5135$  & $1.0260$      & $-0.0051$ & 
      $-0.0693$ & $0.0005$  & $0.0010$ \\
NLL & $-0.3026$  & $1.0081$      & $-0.0048$ &
      $-0.0836$ & $0.0003$  & $0.0009$
\\
\hline
\rule[-2mm]{0mm}{7mm}
 & $C_7^{\rm eff}$ & $C_8^{\rm eff}$ & $C_9$ & $C_{10}$
 & $C_9^{\rm NNLL}$ &  $C_{10}^{\rm NNLL}$ \\
\hline
\rule[-0mm]{0mm}{4mm}
LL  &$-0.3150$ & $-0.1495$ & 
     $2.0072 $ & $0$
 & & \\
NLL & $-0.3094$ & $-0.1695$ & 
      $4.1802$   & $-4.3810$ 
 & \raisebox{2.5mm}[-2.5mm]{$4.2978$} 
 & \raisebox{2.5mm}[-2.5mm]{$-4.4300$} \\
\hline\hline
\end{tabular}
\end{center}
\end{table}

In the QCD factorization formalism the hadronic matrix elements 
are computed in terms of $B$ meson form factors and hadron light-cone 
distribution amplitudes at leading power in a $1/m_b$ expansion. 
Since the semi-leptonic operators ${\cal
  O}_{9,10}$ are bilinear in the quark fields, their 
matrix elements can be expressed directly through $B\to V$ form
factors. The other operators contribute to the decay amplitude only 
through the coupling to a virtual photon, which then decays into the 
lepton pair. We therefore introduce
\begin{eqnarray}
\label{Matrixel}
\langle \gamma^*(q,\mu) V(p',\varepsilon^*)| H^{(i)}_{\rm eff} | \bar B(p)
\rangle &=&  \frac{i g_{\rm em}
  m_b}{4\pi^2}
\Bigg\{2 \,{\cal T}_\perp^{(i)}(q^2) \,\epsilon^{\mu\nu\rho\sigma}
\varepsilon^{\ast}_\nu\, p_\rho p^{\prime}_\sigma\nonumber\\[0.0cm]
&& \hspace*{-4cm}
-2 i\,{\cal T}_\perp^{(i)}(q^2)\left[
E M_B\,\varepsilon^{\ast\mu}-
(\varepsilon^\ast\cdot q)\,p^{\prime\,\mu}\right]
\nonumber\\[0.0cm]
&& \hspace*{-4cm}
-i\,{\cal T}_\parallel^{(i)}(q^2)\,(\varepsilon^\ast\cdot
q)\left[q^\mu-\frac{q^2}{M_B^2}\,(p^\mu+p^{\prime\,\mu})\right]
\Bigg\},
\end{eqnarray}
where $|{V}\rangle$ denotes $|\rho^-\rangle$ ($|K^{*-}\rangle$) for  
$B^-$ meson decay, and $-\sqrt2 |\rho^0\rangle$ or $\sqrt2 |\omega^0\rangle$ 
($|\bar K^{*0}\rangle$) for $\bar B^0$ decay. In the heavy quark
limit this matrix element depends on only two independent functions 
${\cal T}_{a}^{(i)}$ corresponding to a transversely ($a=\perp$) and
longitudinally ($a=\parallel$) polarized $V$. Most of the functions 
${\cal T}_{a}^{(t)}$ can be directly inferred from  the 
calculation for the case $B \to K^*\ell^+\ell^-$~\cite{Beneke:2001at} 
with obvious replacements for $K^* \to \rho(\omega)$. In
Appendix~\ref{app:a} we give the new functions 
${\cal T}_{a}^{(u)}$ and point out the differences 
in ${\cal T}_{a}^{(t)}$ for the $\rho$ $(\omega)$ meson 
compared to $K^*$. The decay amplitudes can be further expressed as
\begin{eqnarray}
\label{calT}
{\cal T}_a^{(i)} 
&=& \xi_a \, C_a^{(i)}  + \frac{\pi^2}{N_c}
\,\frac{f_B f_a}{M_B}\, \Xi_a
\sum_{\pm} \int\frac{d\omega}{\omega}\,\Phi_{B,\,\pm}(\omega)
\int_0^1 \!du\,\, \phi_a(u) \, T_{a,\,\pm}^{(i)}(u,\omega).
\end{eqnarray}
The second term incorporates hard scattering of the spectator
quark. $f_B$ and $\Phi_{B,\pm}$ refer to the $B$ meson decay constant 
and light-cone distribution amplitudes, $f_\parallel\equiv f$, 
$f_\perp$ and $\phi_{a}$ to the corresponding quantities for light 
mesons. Furthermore $\Xi_\perp=1$, $\Xi_\parallel=m_V/E$. The first 
``form factor'' term is expressed in terms of the transverse and 
longitudinal ``soft'' form factors 
$\xi_\perp$ and $\xi_\parallel$, and $C_a$, $T_{a,\pm}$ denote 
perturbative hard scattering kernels. The significance of these terms
will be discussed subsequently, but we mention here 
that in this work we use a definition of $\xi_\parallel$ that differs 
slightly from~\cite{Beneke:2001at,Beneke:2000wa} as explained below. 
In the description of $B\to V\gamma$, there is no advantage in 
using the ``soft'' form factors. In this case we write 
${\cal T}_\perp^{(i)} = T_1(0) \, C_\perp^{(i)\prime}+\ldots$. 
The coefficients $C_\perp^{(i)\prime}$ and $T_{\perp,+}^{(i)\prime}$ 
are related to the unprimed coefficients by (\ref{relation}) 
given below.

\subsection{Input parameters}

\begin{table}[t] 
\centerline{\parbox{14cm}{\caption{\label{tabInput}
Summary of input parameters and estimated uncertainties.}}}
\vspace{-0.1cm}
   \begin{center} 
     \begin{tabular}{| l l| l l |} 
\hline  
\hline 
\rule[-2mm]{0mm}{7mm}
     $\!\!M_W$                           & $80.4$~GeV & 
        $\lambda_{B,+} (1.5{\rm GeV})$ 
           & $0.485 \pm 0.115 \,{\rm GeV}$\\
     $\hat m_t(\hat m_t)$            & $170 \pm 5$~GeV & 
     $\tau_{B^0},\, \tau_{B^\pm}$    & $1.54\,{\rm ps},\, 1.67\,{\rm ps}$ \\
     $\alpha_{\rm em}$               & $ 1/137$ &
        $\lambda$                       & $0.2265^{+0.0025}_{-0.0023}$ \\
     $\Lambda_{\rm QCD}^{(n_f=5)}$   & $220 \pm 40$~MeV &
        $A$                             & $0.801^{+0.029}_{-0.020}$ \\
     $m_{b,\rm PS}(2\,\mbox{GeV})$   & $4.6 \pm 0.1$~GeV &
        $\bar\rho$                      & $0.189^{+0.088}_{-0.070}$ \\
     $m_c$                           & $1.5 \pm 0.2$~GeV &
        $\bar\eta$                      & $0.358^{+0.046}_{-0.042}$\\
     $f_B$                           & $200 \pm 30$~MeV  &
                         & 
\\[0.15cm]
\hline
\rule[-2mm]{0mm}{7mm}
     $\!\!f_{K^*,\perp}$  & $175 \pm 25$~MeV &
        $f_{\rho(\omega),\perp}$  & $150 \pm 25$~MeV \\
     $f_{K^*,\parallel}$             & $218 \pm 4$~MeV & 
        $f_{\rho,\parallel}$, $f_{\omega,\parallel}$ 
                   & $209 \pm 1$~MeV, $187 \pm 3$~MeV \\ 
     $a_1(\bar K^*)_{\perp,\,\parallel}$    &  $0.2 \pm 0.2$ &
        $a_1(\rho,\omega)_{\perp,\,\parallel}$    & $0$ \\ 
     $a_2(\bar K^*)_{\perp,\,\parallel}$    & $0.1 \pm 0.3$  &
        $a_2(\rho)_{\perp,\,\parallel}$,
         $a_2(\omega)_{\perp,\,\parallel}\!\!$  & $0.1 \pm 0.3$, $0.0 \pm
         0.3$  
\\[0.15cm]
\hline
\rule[-2mm]{0mm}{7mm}
     $\!\!M_B\,\xi_{K^*,\parallel}(0)/(2 m_{K^*})\!\!$   & $ 0.47 \pm 0.09$ &
        $M_B\,\xi_{\rho,\parallel}(0)/(2 m_{\rho})$   & $ 0.37 \pm 0.06$ \\
     $\xi_{K^*,\perp}(0)^{\P}$               & $  0.36 \pm 0.07$ &
        $\xi_{\rho,\perp}(0)$                & $ 0.27 \pm 0.05$ 
\\[0.15cm]
\hline\hline
\multicolumn{4}{l}{\footnotesize 
${}^{\P}\,$In Section~\ref{sec31} we determine 
$\xi_{K^*,\perp}(0)=0.26$ from experimental data. This value rather than the 
}\\[-0.1cm]  
\multicolumn{4}{l}{\footnotesize 
\phantom{${}^{\P}\,$}one in the 
Table is then used in the subsequent analysis.}  

       \end{tabular} 
\end{center} 
\end{table} 

A detailed discussion of the input parameters can be found 
in~\cite{Beneke:2001at}. A summary is given in
Table~\ref{tabInput}, where we have taken the  CKM parameters  
from~\cite{Charles:2004jd}. Unless stated otherwise, $m_b$ denotes
the potential-subtracted (PS) heavy quark mass~\cite{Beneke:1998rk} 
(see also the appendix). For the top quark mass we use the 
$\overline{\rm MS}$ definition.

The numerical value of $\lambda_{B,+}$, related to the first inverse moment 
of the $B$ meson light-cone distribution, is taken from the 
QCD sum rule calculation \cite{Braun:2003wx}. The decay constants and 
Gegenbauer moments of the light meson distribution amplitudes follow 
the values given in Table~1 of~\cite{Beneke:2003zv}. We assume large 
uncertainties for these parameters, which cover in particular the 
recent evaluations of the first Gegenbauer moment of the 
kaon~\cite{Khodjamirian:2004ga,Braun:2004vf}\footnote{In our
  convention the momentum fraction used in the light-cone distribution
  amplitudes refers to the outgoing quark. Hence our $a_1(K^*)$ is
  opposite in sign compared to these papers.} 
(renormalization-scale
dependent quantities are evaluated at scale $2$~GeV).  

We find it convenient to slightly change the convention for the 
longitudinal ``soft'' form factor compared 
to~\cite{Beneke:2001at,Beneke:2000wa}. Denoting by a tilde 
the form factors in the old convention, the ``soft'' form factors 
are {\em defined} by the relations 
\begin{eqnarray}
  \xi_\perp(q^2) &=& \tilde \xi_\perp(q^2) = \frac{M_B}{M_B + m_V} \,
  V(q^2),
  \nonumber \\
  \xi_\parallel(q^2) &=& \Delta_\parallel(q^2)\,\tilde\xi_\parallel(q^2) =
   \frac{M_B + m_V}{2E}\,A_1(q^2)-\frac{M_B - m_V}{M_B}\,A_2(q^2),
\label{xiperp}
\end{eqnarray}
where $\Delta_\parallel(q^2)$, given by Eq.~(66)
of~\cite{Beneke:2001at}, determines the perturbative correction to  
the form factor relations between $A_0$, $A_1$ and $A_2$. 
Since $\Delta_\parallel(0)=1$ and $\Delta_\parallel(q^2)=1$ at tree level, the
only difference is a change 
in the $q^2$-dependence of $\xi_\parallel(q^2)$ versus 
$\tilde\xi_\parallel(q^2)$. 

The $q^2$-dependence of the QCD form factors $V$, $A_1$ and $A_2$ has 
been parameterized in the form \cite{Ball:1998kk} 
\begin{equation}
  F(q^2) =  \frac{F(0)}{(1- a_F \, q^2/M_B^2 + b_F \, q^4/M_B^4)}. 
\end{equation}
We use the definitions (\ref{xiperp}) and this equation with 
parameters given in Table~\ref{TabMeson} to determine the 
$q^2$-dependence of $\xi_{\perp}$ and $\xi_{\parallel}$. The
normalization of the longitudinal form factor at $q^2=0$ is  
\begin{equation}
   \xi_\parallel(0) = \frac{2m_V}{M_B}\,A_0(0) =
   \frac{M_B + m_V}{M_B}\,A_1(0)-\frac{M_B - m_V}{M_B}\,A_2(0),
\end{equation}
but we do {\em not} use $V(0)$ to obtain $\xi_\perp(0)$. We use instead 
the relation between $V(0)$ and $T_1(0)$ (at scale 
$\mu=m_b$)~\cite{Beneke:2000wa} 
to write 
\begin{equation}
  T_1(0)= \xi_\perp(0) 
  \left( 1 - \frac{\alpha_s C_F}{4\pi} \right)
     + \frac{\alpha_s(\mu_f) C_F}{4\pi} 
   \, \frac{4 \pi^2 f_B f_\perp}{N_c M_B} \, 
       \lambda_{B,+}^{-1} \, \langle \bar u^{-1} \rangle_\perp
\label{relation}
\end{equation}
with $\mu_f \approx 1.5\,$GeV, and obtain $\xi_\perp(0)$ from  
the value of $T_1(0)$ given in~\cite{Ball:1998kk}. 
This ensures that, since 
the radiative decays $B\to V\gamma$ involve only the tensor form
factor $T_1$ or, alternatively, $\xi_\perp$, 
we obtain identical results for these decays 
independent of whether we express the decay
amplitude in terms of the QCD form factor $T_1$ 
or the ``soft'' form factor. 

Except for the longitudinal decay constant and second 
Gegenbauer moment, the hadronic parameters of
the $\omega$-meson are assumed to be identical to those of the 
$\rho$-meson for lack of more information. As a consequence we do not
have meaningful estimates of the difference between the $\rho^0$ and 
$\omega$ decay rates.

\begin{table}[t]
\centerline{\parbox{14cm}{\caption{\label{TabMeson}
Form factor parameters from QCD sum rules \cite{Ball:1998kk}. 
   The value $V(0)$ is calculated
  from $T_1(0)$ via (\ref{relation}) and (\ref{xiperp}).}}}
\begin{center}
\begin{tabular}{|l||c | c | c || c | c | c || c | c | c|| c|}
\hline\hline
  meson & $V(0)$ & $a_V$ & $b_V$ & $A_1(0)$ & $a_1$ & $b_1$ & 
$A_2(0)$ & $a_2$ & $b_2$ & $T_1(0)$\\
\hline\hline
  $\rho$   & $0.313$ & $1.37$ & $0.315$ & $0.261$ & $0.29$ & $-0.415$
  & $0.223$ & $0.93$ & $-0.092$ & $0.285$\\
  $K^*$    & $0.424$ & $1.55$ & $0.575$ & $0.337$ & $0.60$ & $-0.023$
  & $0.283$ & $1.18$ & $\phantom{-}0.281$ & $0.379$\\
\hline\hline
\end{tabular}
\end{center}
\end{table}

\subsection{Decay rates and distributions}

The decay amplitudes for $B\to V\gamma$ and 
$B\to V\ell^+\ell^-$ can be written in terms of~\cite{Beneke:2001at}
\begin{eqnarray}
{\cal C}_7^{(i)} &\equiv& \frac{{\cal T}_{\perp}^{(i)}(0)}{T_1(0)} 
= \delta^{it}\,C_7^{\rm eff} + \ldots,
\nonumber \\
{\cal C}_{9,\,\perp}^{(i)}(q^2) &\equiv& \delta^{it}\,C_9+
\frac{2 m_b M_B}{q^2}\,\frac{{\cal T}_{\perp}^{(i)}(q^2)}{\xi_\perp(q^2)},
\nonumber \\
{\cal C}_{9,\,\parallel}^{(i)}(q^2) &\equiv& \delta^{it} \,C_9 
 -\frac{2 m_b}{M_B}\,
 \frac{{\cal T}_{\parallel}^{(i)}(q^2)}{\xi_\parallel(q^2)},
\label{CalC9L}
\end{eqnarray}
where $i=t,u$ refers to the two different CKM factors.\footnote{
\label{foot1} There 
is a factor of $\Delta_\parallel(q^2)$ missing in Eq.~(41) of 
\cite{Beneke:2001at}. With the new definition of the $\xi_\parallel$ 
this factor is absent from the definition of 
${\cal C}_{9,\,\parallel}^{(i)}$. Furthermore, we now define  
${\cal C}_7^{(i)}$ by dividing through the QCD tensor form factor (at scale
$\mu=m_b$) rather than $\xi_\perp$. Hence ${\cal C}_7^{(i)}$ is 
expressed in terms of the coefficients $C_\perp^{(i)\prime}$ and 
$T_{\perp,+}^{(i)\prime}$ defined at the end of 
Section~\ref{sec:formalism} (see also the appendix).}

With these definitions, we obtain the decay rate for 
$\bar B\to \rho(\omega)\gamma$ in the form 
\begin{eqnarray}
\label{Gamma}
\Gamma(\bar B\to \rho(\omega)\gamma) &=& \frac{G_F^2}{8\pi^3}\,
M_B^3\, S\left(1-\frac{m_V^2}{M_B^2}\right)^{\!3}
\frac{\alpha_{\rm em}}{4\pi}\,m_{b}^2\,T_1(0)^2 \CR
&&\times\left\{|\lambda_t^{(d)}{\cal C}_7^{(t)}|^2
+|\lambda_u^{(d)}{\cal C}_7^{(u)}|^2
-2|\lambda_u^{(d)}\lambda_t^{(d)}|
{\mbox Re}\left(e^{i\alpha}{\cal C}_7^{(u)}{\cal C}_7^{(t)*}\right)\right\},
\end{eqnarray}
with  $S=1$ for $\rho^-$, 
and $S=1/2$ for $\rho^0$ and $\omega$. 
$m_V$ denotes the mass of the light meson, which we include in the phase space
factor, while in general $m_V^2$ terms are neglected. The dominant
term is $|\lambda_t^{(d)}{\cal C}_7^{(t)}|^{\,2} \sim |V_{td}|^2$, 
but the interference term is non-negligible and can be the source of
interesting CP-violating and isospin-breaking effects. The
CP-conjugate $B$ decay follows from (\ref{Gamma}) 
by the replacement $\alpha\to-\alpha$. 

The decay $\bar B\to \rho(\omega) \ell^+ \ell^-$ has a richer kinematic
structure. Defining $q^2$, the invariant mass of the lepton pair, and
$\theta$, the angle between the positively charged lepton and the $\bar
B$ meson in the center-of-mass frame of the lepton pair, and summing 
over final state polarizations, the decay information is contained 
in the double differential distributions 
\begin{eqnarray}
\label{dGamma}
\frac{d^2\Gamma}{dq^2 d\!\cos\theta} 
&=& \frac{G_F^2}
{128\pi^3}\,M_B^3\, S\,\lambda(q^2,m_V^2)^3
\left(\frac{\alpha_{\rm em}}{4\pi}\right)^{\!2}\times\Bigg\{ 
\nonumber\\
&&\hspace{-2.5em}
(1+\cos^2\theta)\,\frac{2 q^2}{M_B^2}\,\xi_\perp(q^2)^2 
\Bigg[\sum_{q=u,t}|\lambda_q^{(d)}|^2\left(
|{\cal C}_{9,\,\perp}^{(q)}(q^2)|^2+\delta_{qt}\,C_{10}^2\right)\CR
&&\hspace{11em}
-2|\lambda_u^{(d)}\lambda_t^{(d)}|
\,{\mbox Re} \left( e^{i\alpha}\,
{\cal C}_{9,\,\perp}^{(u)}(q^2)\,{\cal C}_{9,\,\perp}^{(t)*}(q^2)\right)\Bigg]
\nonumber\\
&& \hspace{-2.5em}
+\,(1-\cos^2\theta)\,
\left(\frac{E\,\xi_\parallel(q^2)}{m_V}\right)^{\!2}\,
\Bigg[\sum_{q=u,t}|\lambda_q^{(d)}|^2\left(
\,|{\cal C}_{9,\,\parallel}^{(q)}(q^2)|^2+\delta_{qt}\,C_{10}^2\right)
\label{gammadist}\\
&&\hspace{11em}
-2|\lambda_u^{(d)}\lambda_t^{(d)}|
\,{\mbox Re} \left( e^{i\alpha}\,
{\cal C}_{9,\,\parallel}^{(u)}(q^2)\,{\cal C}_{9,\,\parallel}^{(t)*}(q^2)
\right)\Bigg]
\nonumber\\
&&\hspace{-2.5em}
-\cos\theta\,\frac{8 q^2}{M_B^2}\,\xi_\perp(q^2)^2\,C_{10}
\left[|\lambda_t^{(d)}|^2 \,
\mbox{Re}\left({\cal C}_{9,\,\perp}^{(t)}(q^2)\right)
-|\lambda_u^{(d)}\lambda_t^{(d)}|
\,{\mbox Re}\left( e^{i\alpha}\,{\cal C}_{9,\,\perp}^{(u)}(q^2)
\right)\right]\Bigg\} \nonumber,
\end{eqnarray}
where 
\begin{equation}
\lambda(q^2,m_V^2) = \Bigg[\left(1-\frac{q^2}{M_B^2}\right)^2-
    \frac{2 m_V^2}{M_B^2}
\left(1+\frac{q^2}{M_B^2}\right)+\frac{m_V^4}{M_B^4}\Bigg]^{\!1/2},
\end{equation}
The lepton mass is set to zero, so this result applies to
$\ell=e,\mu$.  The
CP-conjugate $B$ decay follows from (\ref{gammadist}) 
by the replacement 
$\alpha\to-\alpha$. The terms with angular dependence $(1\pm \cos^2\theta)$ 
correspond to the decay into transversely and longitudinally
polarized $\rho(\omega)$'s, respectively. The term proportional 
to $\cos\theta$ generates a forward-backward asymmetry with respect 
to the plane perpendicular to the lepton momentum in the
center-of-mass frame of the lepton pair.

\subsection{Overview of amplitudes}

\begin{table}[t]
\caption{\label{tabCalC7}
Breakdown of the decay amplitudes ${\cal C}_7^{(i)}$ for $B \to
\rho\gamma$ and $\omega\gamma$. }
\begin{center}
\begin{tabular}{|l|| c | c |}
\hline\hline
 & ${\cal C}_7^{(t)}$ & ${\cal C}_7^{(u)}$
\\
\hline\hline
&& \\[-0.45cm]
$C_7^{\rm eff}$ & $-0.309$ & 0 
\\
$C^{(1)\prime}$ & $-0.060-0.013 i$ & $\phantom{+}0.063+0.055i$
\\
\hline
&& \\[-0.45cm]
$T^{(0)}$ (annih.) & 0 & 0 
\\
$T^{(1)\prime} \, \begin{array}{l}
 \;\:(\rho)\\ \;\:(\omega)\\
\end{array}$&
$\begin{array}{r} -0.009-0.012 i \\ -0.009-0.013 i   \end{array}$ &
$\begin{array}{r} -0.031-0.012 i\\  -0.029-0.013 i   \end{array}$
\\
\hline
&& \\[-0.45cm]
$\alpha_s^0/m_b$ (annih.)&
$\begin{array}{rl} \phantom{+}0.009 & \;\;(\rho^-) \\ 
-0.005 & \;\;(\rho^0) \\ -0.004 & \;\;(\omega) \end{array} $ &
$\begin{array}{rl} -0.125  & \;\;(\rho^-)\\ -0.012  & \;\;(\rho^0)\\ 
\phantom{+}0.011 & \;\;(\omega)\end{array}$ 
\\
\hline
&& \\[-0.45cm]
$\begin{array}{rl}
&  \;\;(\rho^-)\\
\alpha_s^1/m_b & \;\;(\rho^0)\\
& \;\;(\omega)
\end{array}$
&$\begin{array}{r} \phantom{+}0.001+0.000i \\ \phantom{+}0.000-0.001i 
\\ \phantom{+}0.001-0.001i \end{array}$ & 
$\begin{array}{r} \phantom{-}0.000-0.004i \\ \phantom{+}0.000+0.002i \\
\phantom{-}0.000+0.002i \end{array}$ 
\\
\hline\hline
&& \\[-0.45cm]
$\begin{array}{rl}
&  \;\;(\rho^-)\\
\mbox{sum} & \;\;(\rho^0)\\
& \;\;(\omega)
\end{array}$
&$\begin{array}{r} -0.369-0.024i \\ -0.383-0.026i \\ 
-0.383-0.027i \end{array}$ & 
$\begin{array}{r} -0.093+0.039i \\  \phantom{+}0.019+0.045i \\
\phantom{-}0.045+0.042i \end{array}$ \\
\hline\hline
\end{tabular}
\end{center}
\end{table}

We briefly discuss the qualitative features of the decay amplitudes
from which the main characteristics of the observables that we analyze
in the following section can be deduced. Beginning with 
$B\to V\gamma$ we see from Table~\ref{tabCalC7} that the process 
is dominated by the short-distance electromagnetic penguin amplitude 
proportional to $C_7^{\rm eff}$. However, there is an important 
radiative correction $C^{(1)\prime}$ 
to the quark matrix element $\langle \gamma 
d | H _{\rm eff}^{(t,u)}|b\rangle$ \cite{Greub:1996tg}, 
as well as a sizeable correction from spectator scattering 
\cite{Beneke:2001at,Bosch:2001gv} specific to the exclusive decays. 
It is worth noting that these two corrections add constructively 
in the top-sector of the effective Hamiltonian (${\cal C}_7^{(t)}$), 
but tend to cancel in the up-sector. The most distinctive feature 
of exclusive $b\to d\gamma$ transitions is the large weak annihilation
contribution to $B^\pm\to\rho^\pm\gamma$, which has been discussed 
extensively~\cite{Bosch:2001gv,Ali:2001ez,Khodjamirian:1995uc,Ali:1995uy,Grinstein:2000pc}. Although formally suppressed by a factor 
of $1/m_b$, the annihilation amplitude is the largest contribution 
to ${\cal C}_7^{(u)}$ for $B^\pm\to\rho^\pm\gamma$, because it is 
enhanced by a Wilson coefficient ten times larger than  $C_7^{\rm
  eff}$. The leading annihilation amplitude is short-distance 
dominated and can be computed in the heavy quark limit. However, as 
discussed in the Appendix, there is a large theoretical uncertainty 
associated with this computation due to both parameter uncertainties 
($f_B, \lambda_{B+}$) and power corrections. The remaining 
power-suppressed amplitudes listed in Table~\ref{tabCalC7} 
are small effects and relevant only to the extent that they provide 
the leading source of difference between $\rho^0$ and $\omega$, or, 
in the case of $b\to s $ transitions $K^{*+}$ and  
$K^{*0}$ \cite{Kagan:2001zk}.  
The Table shows that 
the difference of $\rho^0$ and $\omega$ is negligible, but it should
be remembered that the hadronic parameters of the $\omega$ have 
been set equal to those of $\rho^0$ in the absence of better information.

From a phenomenological point of view, the complete information 
about the four different $B\to \rho\gamma$ decays is encoded 
in the parameters $|({\cal C}^{(t)}_7)^{\rho^0}|\simeq 0.38$, 
\begin{equation}
  \frac{({\cal C}_7^{(t)})^{\rho^+}}{({\cal C}_7^{(t)})^{\rho^0}} -1
  \equiv \frac{\delta_+ + i \eta_+}{2} \simeq -0.04 + 0.00 i,
\label{abbrev1}
\end{equation}
and 
\begin{eqnarray}
  \frac{({\cal C}_7^{(u)})^{\rho^0}}{({\cal C}_7^{(t)})^{\rho^0}}
  \equiv \epsilon_0 \, e^{i\theta_0} \simeq - 0.06 - 0.11 i,
  \qquad
  \frac{({\cal C}_7^{(u)})^{\rho^+}}{({\cal C}_7^{(t)})^{\rho^+}}
  \equiv \epsilon_+ \, e^{i\theta_+} \simeq 0.24 - 0.12 i.
\label{abbrev2}
\end{eqnarray}
The discussion of the theoretical errors of these parameters is 
deferred to the following sections. The decay amplitudes are then 
proportional to 
$1-R_{ut} \epsilon_i e^{i (\theta_i\pm\alpha)}$ ($i=0,+$),
where 
\begin{eqnarray}
   \lambda_u^{(d)}/\lambda_t^{(d)} = - R_{ut} \, e^{i\alpha}
\label{Rut}
\end{eqnarray}
with $R_{ut}\simeq 0.46$, and $\alpha \simeq 94^\circ$ in the Standard
Model, and the upper (lower) sign refers to the decay of a $\bar B$ ($B$) 
meson.

The amplitude structure of $B\to V\ell^+\ell^-$ is somewhat 
more complicated due to the presence of the axial-vector
short-distance contribution proportional to $C_{10}$,  
the duplication of amplitudes for transverse and longitudinal
polarization of the vector meson, and the $q^2$-dependence of 
these amplitudes. For very small $q^2$ the amplitude 
is dominated by the photon pole and exhibits a behaviour 
qualitatively very similar to that discussed above for a real photon. 
In most of the region, in which our theoretical treatment is 
applicable ($q^2< 7\,\mbox{GeV}^2$), $C_{10}$ and 
the longitudinal amplitude determine the decay characteristics 
with the exception of the forward-backward asymmetry. The  
latter is directly proportional to the expression in square 
brackets in the last line of (\ref{gammadist}), and is 
therefore sensitive to the real part of ${\cal C}_{9,\perp}^{(t)}$ 
and $e^{\pm i\alpha}\,{\cal C}_{9,\perp}^{(u)}$, which is expected 
to go through zero in the $q^2$-range of interest. 
Table~\ref{tabCalC9} gives the values for ${\cal
  C}_{9,a}^{(t,u)}$ ($a=\perp,\parallel$) at $q^2=5\,\mbox{GeV}^2$ 
for the decays to the $\rho$ meson. We note again that the 
two-loop correction $C^{(1)}$ to the quark matrix element 
$\langle  \gamma^* d | H _{\rm eff}^{(t,u)}|b\rangle$ is rather large, 
in fact as large as the contributions from four-quark operators 
at leading order in $\alpha_s$ contained in the functions 
$Y^{(i)}(q^2)$. This is true in particular for the amplitude in the up-sector, 
which was calculated only recently \cite{Seidel:2004jh} 
(see also \cite{Asatrian:2003vq}). Similarly to the radiative decay 
the power-suppressed weak annihilation amplitude is the largest
remaining term in the transverse amplitude of the decay to 
$\rho^\pm$. Turning to the longitudinal amplitudes  
${\cal C}_{9,\parallel}^{(t,u)}$, we should recall that we do not 
include $1/m_b$-corrections in this case. The reason for this is 
that the important isospin-breaking effects exist already 
at leading power as is evident from the entry for $T^{(0)}$ (annih.) 
in Table~\ref{tabCalC9}. This comes from a weak annihilation 
contribution to $B\to\rho \, \ell^+\ell^-$, which is not suppressed 
by factors of $\alpha_s$ or $1/m_b$ relative to the short-distance 
amplitude~\cite{Beneke:2001at}, and which now also has a non-vanishing (and 
rather uncertain) absorptive part.
This leads to a large difference of 
the longitudinal amplitude in the up-sector between the charged and 
neutral $\rho$ final state similar to the situation for 
${\cal C}_7^{(u)}$ (now including the absorptive part). 
The amplitude in the up-sector is correspondingly uncertain (see 
the discussion in the appendix). 

\begin{table}[t]
\caption{\label{tabCalC9}
Breakdown of the decay amplitudes 
${\cal C}_{9,a}^{(t,u)}(q^2~=~5~{\rm GeV}^2)$ for $B \to
\rho \,\ell^+\ell^-$. We define $a C_7^{\rm eff} \equiv (2 m_b M_B/q^2) \, 
C_7^{\rm eff}$ for ${\cal C}_{9,\perp}^{(t)}$ and 
 $a C_7^{\rm eff} \equiv (2 m_b/M_B)\,  
C_7^{\rm eff}$ for ${\cal C}_{9,\parallel}^{(t)}$.}
\begin{center}
\begin{tabular}{|l|| c | c | c | c |}
\hline\hline
&&&& \\[-0.45cm]
 & ${\cal C}_{9,\perp}^{(t)}$ & ${\cal C}_{9,\perp}^{(u)}$
 & ${\cal C}_{9,\parallel}^{(t)}$ & ${\cal C}_{9,\parallel}^{(u)}$
\\
\hline\hline
&& \\[-0.45cm]
$C_9$ &  $ \phantom{+}4.30$ & 0 & $ \phantom{+}4.30$ & 0 
\\
$Y(q^2)$ & $\phantom{+}0.51 + 0.06 i$ & $-0.07 - 0.84 i$ & 
           $\phantom{+}0.51 + 0.06 i$ & $-0.07 - 0.84 i$ 
\\
$a C_7^{\rm eff}$ & $-3.01$ & 0 & $-0.54$ & 0 
\\
$C^{(1)}$ &$-0.82 - 0.14 i$ & $\phantom{+}0.23 + 1.27 i$ & 
           $-0.39 - 0.01 i$ & $-0.07 +0.72 i$
\\
\hline
&&&& \\[-0.45cm]
$\begin{array}{ll}
T^{(0)} &  \;(\rho^-)\\
(\mbox{annih.}) & \;(\rho^0)\\
\end{array}$ & 0 & 0 &
$
\begin{array}{r}
  \phantom{+}0.04 - 0.03 i     \\ -0.02 + 0.01 i \end{array} 
$ & 
$ \begin{array}{r}
   \phantom{+}1.31 - 0.86 i    \\ \phantom{+}0.13 - 0.09 i\end{array}$
\\
$\begin{array}{ll}
T^{(1)} &  \;(\rho^-)\\
\phantom{(\mbox{annih.})} & \;(\rho^0)\\
\end{array}$
& $-0.21 - 0.15 i$ &  $-0.09 - 0.15 i$ &
$ \begin{array}{r}
\phantom{-}0.05 - 0.04 i  \\ 
\phantom{-}0.02 - 0.05 i
\end{array}$ & 
$ \begin{array}{r}
-0.08 - 0.07 i \\
-0.09 - 0.04 i 
\end{array} $
\\
\hline
&&&& \\[-0.45cm]
$\alpha_s^0/m_b \, \begin{array}{l}
 \;\:(\rho^-)\\ \;\:(\rho^0)\\
\end{array}$&
$\begin{array}{r} \phantom{+}0.07 - 0.03 i \\ -0.04 + 0.02 i   \end{array}$ & 
$\begin{array}{r} \phantom{+}0.39 - 1.06 i \\  \phantom{+}0.04 - 0.11
  i   \end{array}$ &
--- & ---
\\
\hline
&&&& \\[-0.45cm]
$\alpha_s^1/m_b \, \begin{array}{l}
 \;\;(\rho^-)\\
\;\;(\rho^0)\\
\end{array}$&
$\begin{array}{r} \phantom{-} 0.03 + 0.00 i \\  
-0.02 + 0.00 i  \end{array} $ & 
$\begin{array}{r}  \phantom{+}0.00 - 0.03 i \\ 
\phantom{+} 0.00 + 0.02 i  \end{array} $ & 
--- & ---
\\
\hline
&&&& \\[-0.45cm]
$\mbox{sum} \, \begin{array}{l}
 \;\;(\rho^-)\\
\;\;(\rho^0)\\
\end{array}$ &
$\begin{array}{r} \phantom{-} 0.87 - 0.25 i \\  
\phantom{+}0.72 - 0.21 i  \end{array} $ & 
$\begin{array}{r}  \phantom{+}0.46 - 0.81 i \\ 
\phantom{+} 0.10 + 0.20 i  \end{array} $ & 
$\begin{array}{r} \phantom{-} 3.97 - 0.00 i \\  
\phantom{+} 3.88 + 0.02 i  \end{array} $ & 
$\begin{array}{r}  \phantom{+}1.08 - 1.05 i \\ 
-0.11 - 0.25 i  \end{array} $
\\
\hline\hline
\end{tabular}
\end{center}
\end{table}

In the following sections we perform a detailed analysis of the
observables of interest, including theoretical error estimates not
represented in Tables~\ref{tabCalC7}~and~\ref{tabCalC9}. The 
tables should provide useful information to understand the 
numerics and parameter dependences of the various observables.

\section{Phenomenological analysis}
\label{sec:pheno}

In this section we discuss several observables related to 
rare radiative
$B$ decays which are currently studied at $B$ factories, or
will be measured at future high-luminosity experiments. 
The $B \to V\gamma$ decays have been recently studied within the QCD 
factorization framework by Ali et al.~\cite{Ali:2004hn}, 
and by Bosch and Buchalla~\cite{Bosch:2004nd}, and
in most aspects our analysis leads
to similar results. In the case of $B \to K^*\ell^+\ell¯$ decays we perform
an update of our previous work \cite{Beneke:2001at}. Finally, we extend
the discussion to $B \to \rho(\omega)\ell^+\ell^-$ decays. Here
the inclusion of the recently calculated two-loop 
correction to the $b \to d\gamma^*$ vertex from light-quark loops
\cite{Seidel:2004jh,Asatrian:2003vq} turns out to have a significant numerical
impact on the various decay asymmetries. In the following discussion 
we focus on the sensitivity of the radiative decays  
to the theoretical input parameters, including the CKM 
elements, short-distance Wilson coefficients, and hadronic parameters.

\subsection{$B \to V\gamma$ branching fractions}
\label{sec31}

\begin{table}[t] 
\centerline{\parbox{14cm}{\caption{\label{tabExp}
Experimental results for the CP-averaged $B \to V\gamma$ branching ratios 
(in units of $10^{-6}$).}}}
\vspace{0.1cm}
\begin{center} 
\renewcommand{\arraystretch}{1.3}
\begin{tabular}{| l | c | c | c | c |} 
\hline  
\hline 
\rule[-2mm]{0mm}{7mm}
& BABAR~\cite{Aubert:2004fq,Aubert:2004te} 
& BELLE~\cite{Abe:2004kq,Nakao:2004th} 
& CLEO~\cite{Coan:1999kh}
& Average \\\hline
%
%
Br($B^0\to K^{*0}\gamma$) &
$39.2\pm 2.0\pm 2.4$ &
$40.1\pm 2.1 \pm 1.7 $ & 
$45.5^{+7.2}_{-6.8}\pm 3.4$ & 
$40.1\pm 2.0 $\\
%
%
Br($B^+\to K^{*+}\gamma$) &
$38.7\pm 2.8\pm 2.6$ &
$42.5\pm 3.1\pm 2.4$ & 
$37.6^{+8.9}_{-8.3}\pm 2.8$ & 
$40.3\pm 2.6$
\\[0.05cm]
\hline\hline
%
%
Br($B^0\to \omega\gamma$) &
$< 1.0$ &
$< 0.8$ & 
$< 9.2 $ &
$< 0.8 $\\
%
%
%
Br($B^0\to \rho^{0}\gamma$) &
$< 0.4$ &
$< 0.8$ & 
$< 17 $ &
$< 0.4$\\
%
Br($B^+\to \rho^{+}\gamma$) &
$< 1.8$ &
$< 2.2$ &
$< 13 $ &
$< 1.8$ \\
\hline\hline
\end{tabular} 
\end{center} 
\end{table} 

We first consider the $B \to K^*\gamma$ decays. Using
the central values and theoretical uncertainties for the input
parameters in Table~\ref{tabInput}, we obtain for
the branching fractions 
\begin{eqnarray}
  {\rm Br}(B^0 \to K^{*0}\gamma) &=& 
  \left(\frac{T_1^{K^*}(0)}{0.38} \right)^{\!2} \ 
  (7.4 \, {}^{+0.6}_{-0.5} \big|_{V_{ts}} 
       {}^{+0.7}_{-0.7} \big|_{\rm had}) \cdot 10^{-5},
\\[0.2em] 
  {\rm Br}(B^+ \to K^{*+}\gamma) &=& 
  \left(\frac{T_1^{K^*}(0)}{0.38} \right)^{\!2} \ 
  (7.4 \, {}^{+0.6}_{-0.5} \big|_{V_{ts}} 
       {}^{+0.6}_{-0.7} \big|_{\rm had}) \cdot 10^{-5} .       
\end{eqnarray}
The main theoretical uncertainty comes from
the $B \to K^*$ tensor form factor, the other hadronic and CKM
parameter uncertainties being small as 
indicated.\footnote{The theoretical error is computed from the 
parameter uncertainties given in Table~\ref{tabInput}, including the 
renormalization scale dependence, but no attempt is made to quantify
the error from higher-order perturbative corrections and power 
corrections. This should always be kept in mind in the interpretation
of theoretical errors.} Comparison with 
Table~\ref{tabExp} shows that with the form factor from 
light-cone sum rules, the theoretical value is too high compared to 
data. Since the short-distance
physics explored in inclusive $b \to s\gamma$ decays is well in line
with the standard model expectation, we must conclude that 
either there are large power corrections not included in the
computation, or the tensor form factor is smaller than 0.38. 
The existence of large power corrections beyond the known annihilation
terms would invalidate the basic assumption of the factorization 
approach that the heavy quark expansion provides a sensible
approximation. In the absence
of any direct determinations of the form factor, we consider 
the reduction of the form factor a viable option, and use 
the branching fraction data to obtain~\cite{Beneke:2001at}
\beq
  T_1^{K^*}(0) \big|_{\rm exp} = 0.28 \pm 0.02.  
\label{T1Kexp}
\eeq
This number will be taken as a reference value in the following. 

The uncertainty in the form factors is also a major obstruction to 
a clean interpretation of the $B \to \rho\gamma$ decays. The 
situation for $K^*$ may suggest that the $B\to\rho$ tensor form factor
should be reduced in equal proportion, in which case we obtain 
for the branching fractions averaged over $B$ decay and the 
corresponding CP-conjugate $\bar B$ decay  
\begin{eqnarray}
  {\rm Br}(B^0 \to \rho^0\gamma) &=&  
  \left( \frac{|V_{td}|}{8.25 \cdot 10^{-3}} \right)^{\!2}
  \left( \frac{T_1^\rho(0)}{0.21} \right)^{\!2} \,
        (5.0 {}^{+0.5}_{-0.5}) \cdot 10^{-7} ,
\label{Brho0BR}
\\[0.2em]
   {\rm Br}(B^+ \to \rho^+\gamma) &=&  
  \left( \frac{|V_{td}|}{8.25 \cdot 10^{-3}} \right)^{\!2}
  \left( \frac{T_1^\rho(0)}{0.21} \right)^{\!2} \,
        (10.3 {}^{+1.5}_{-1.2}) \cdot 10^{-7} .
\end{eqnarray}
On the other hand, if the origin of the large $B\to K^*$ form factor
in QCD sum rules is a misestimate of SU(3) breaking effects, 
$T_1^\rho(0)\simeq 0.29$ is not excluded and the central values of the 
branching fractions are rescaled by almost a factor of 2. 
The branching fraction for $B^0\to\omega\gamma$ equals 
${\rm Br}(B^0 \to \rho^0\gamma)$ within the given accuracy, and 
under the assumption that the form factors are the same. Comparison 
with the experimental limits on exclusive $b\to d\gamma$ transitions 
in Table~\ref{tabExp} favours the scenario with a small 
form factor when the standard range of $|V_{td}|$ is assumed. 
In the following, however, we continue to use the default value 
$T_1^\rho(0)=0.29\pm 0.04$.

It is often stated that the ratios of $B\to \rho\gamma$ to 
$B\to K^*\gamma$ branching fractions are better suited for a
determination of $|V_{td}|$. This is based on the assumption 
that the ratio $F=T_1^\rho(0)/T_1^{K^*}(0)$ is better known 
than the form factors themselves. Unfortunately, given that we 
do not know for certain whether the current estimates of the form factors 
are affected by a normalization or SU(3) breaking problem, we 
must assume $0.75<F<1.05$ at least.\footnote{The lower limit corresponds to the
  assumption that QCD sum rules predict the ratio of form factors
  correctly. The upper limit assumes that $T_1^\rho(0)$ is predicted
  correctly and $T_1^{K^*}(0)$ is determined by data which yields 
  (\ref{T1Kexp}).}
The ratios of CP-averaged branching fractions are determined by 
\begin{eqnarray}
  \frac{{\rm Br}(B^0 \to \rho^0\gamma)}
       {{\rm Br}(B^0\to  K^{*0}\gamma)}
&=& \frac{1}{2} F^2  \left| \frac{V_{td}}{V_{ts}} \right|^2 
  \left\{ 1 - 2 R_{ut} \, \epsilon_0 \cos\alpha \, \cos\theta_0
   + R_{ut}^2 \, \epsilon_0^2 \right\}
   \nonumber\\
&&\hspace*{-2cm} 
=\, \frac{1}{2} F^2  \left| \frac{V_{td}}{V_{ts}} \right|^2 
  \left\{ 1 - 2 R_{ut}\cos\alpha \,[-0.06^{+0.06}_{-0.06}]
   + R_{ut}^2 \, [0.02^{+0.02}_{-0.01}] \right\}
\label{br_neutral}
\end{eqnarray}
and
\begin{eqnarray}
  \frac{{\rm Br}(B^+ \to \rho^+\gamma)}
       {{\rm Br}(B^+\to  K^{*+}\gamma)}
&=& F^2 \left| \frac{V_{td}}{V_{ts}} \right|^2 
  \left\{ 1 - 2 R_{ut} \, \epsilon_+ \cos\alpha \, \cos\theta_+
   + R_{ut}^2 \, \epsilon_+^2 \right\}
   \nonumber\\
&&\hspace*{-2cm} 
=\, F^2 \left| \frac{V_{td}}{V_{ts}} \right|^2 
 \left\{ 1 - 2 R_{ut} \cos\alpha \,[0.24^{+0.18}_{-0.18}] 
   + R_{ut}^2 \, [0.07^{+0.12}_{-0.07}]  \right\},
\label{br_charged}
\end{eqnarray}
where a small phase space correction 
$(M_B^2-m_\rho^2)^3/(M_B^2-m_{K^*}^2)^3 \simeq 1.02$, and 
a SU(3) breaking correction from 
$ |{\cal C}_7^t|^{\rho}/|{\cal C}_7^t|^{K^{*}} -1 =\pm 0.02$ 
can be safely neglected.\footnote{With $T_1^\rho(0)=0.21$ instead of 
0.29, the central values of the numerical entries in
(\ref{br_neutral}) change from $(-0.06,0.02)$ to $(-0.01,0.01)$ and 
in (\ref{br_neutral}) from $(0.24,0.07)$ to $(0.40,0.18)$.} 
The global CKM fit \cite{Charles:2004jd} 
returns $R_{ut}=0.46\pm 0.06$ and $\cos\alpha=-0.07\pm 0.20$, 
so the interference term is expected to be suppressed. 
Since the value of $\epsilon_0 \cos\theta_0$
is rather small due to a partial cancellation of the perturbative 
corrections $C^{(1)}$ and $T^{(1)}$ (see Table~\ref{tabCalC7}), 
the $B^0 \to \rho^0\gamma$ (and to a slightly lesser 
extent also $B^0\to\omega\gamma$) decay is 
rather insensitive to $R_{ut}$ and $\alpha$. The neutral ratio 
thus provides a direct constraint on $|V_{td}/V_{ts}|$ for
a given value of the ratio of tensor form factors 
$F=T_1^\rho(0)/T_1^{K^*}(0)$. 
For the charged ratio  $\epsilon_+ \cos\theta_+$ is dominated 
by the weak annihilation contribution, to which we assign 
a 50\% error (see the discussion in the appendix). This error is 
by far the largest uncertainty in $B^+\to\rho^+\gamma$, and 
makes the charged ratio somewhat less useful to
constrain  $|V_{td}/V_{ts}|$. 

At this point it is appropriate to compare our results 
$\epsilon_0\cos\theta_0=-0.06\pm 0.06$ and 
$\epsilon_+\cos\theta_+=0.24\pm 0.18$ to 
\cite{Ali:2004hn,Bosch:2004nd}. Bosch and Buchalla employ a formalism
identical to ours and give $0.0\pm 0.1$ and $0.4\pm 0.4$ for these 
parameters. We find agreement with their central values, 
when we adjust the parameter $\lambda_{B,+}$ to the smaller 
value 350 MeV used by them. Ali et al. do not give 
$\epsilon_0\cos\theta_0$ and 
$\epsilon_+\cos\theta_+$ explicitly, but the 
curly brackets in (\ref{br_neutral},\ref{br_charged}) for 
which we obtain $1.00\pm 0.01$ and $1.03\pm 0.06$ correspond  
to their $1+\Delta R = 1.09\pm 0.07$ and $1.12\pm 0.10$.

\subsection{$B \to V\ell^+\ell^-$ branching fractions}

The differential branching fractions for $B \to K^*\ell^+\ell^-$
decays have been discussed in detail in our previous analysis
\cite{Beneke:2001at}. The present work
includes changes of input parameters (see Table~\ref{tabInput}),
the now available NNLO anomalous dimensions relevant
for the Wilson coefficient $C_9$ \cite{Gambino:2003zm,Gorbahn:2004my}, 
as well as a modification of the treatment of form factors. 
In particular, the transverse form factor is now taken from 
data via (\ref{T1Kexp}), and we no longer use the simple $1/E^n$ 
($n=2,3$)  model for the $q^2$-dependence of the form factors. The combined
effect of these changes is to decrease the branching fraction by 
about 30\% relative to~\cite{Beneke:2001at}, almost exclusively due to 
the change in the form factor input. 

The $q^2$-spectrum rises sharply for small $q^2$, where it is
dominated by the photon pole and possibly ``contaminated'' by hadronic
resonances. On the other hand the factorization 
approach is valid only for $q^2 \leq 7$~GeV$^2$, away from the 
charm threshold. We therefore advocate that measurements of exclusive 
branching fractions be compared to the integral 
\begin{eqnarray}
   \int\limits_{1\,{\rm GeV}^2}^{6\,{\rm GeV}^2} \!\!dq^2 \, 
   \frac{d{\rm Br}(B^+\to K^{*+}\ell^+\ell^-)}{dq^2}
 &=& 
\left(\frac{A_0^{K^*}(4\,\mbox{GeV}^2)}{0.66} \right)^{\!2} 
      \, (3.33^{+0.40}_{-0.31}) \cdot 10^{-7}.
\end{eqnarray}
of the CP-averaged spectrum. 
The corresponding integral for the neutral $B$ meson decay is about 10\% 
smaller. The largest hadronic uncertainty arises from the 
longitudinal form factor, which has therefore been scaled out, 
such that only the residual $A_0$-dependence is included in the 
error.

Recently the $B$ factories have presented their first results for 
(partially) integrated $B \to K^*\ell^+\ell^-$ decay rates
\cite{Aubert:2003cm,Ishikawa:2003cp}. Of particular interest 
to us is the second bin of the Belle measurement \cite{Abe:2004ir}, 
which translates into
\begin{eqnarray}
   \int\limits_{4\,{\rm GeV}^2}^{8\,{\rm GeV}^2} \!\!dq^2 \, 
   \frac{d{\rm Br}(B\to K^*\ell^+\ell^-)}{dq^2}
 &=& (4.8 \, {}^{+1.4}_{-1.2}|_{\rm stat.} \,  
             \pm 0.3|_{\rm syst.} \,
             \pm 0.3|_{\rm model}  ) \cdot 10^{-7},
\label{exp:integrate}
\end{eqnarray}
which we interpret as an average of the charged and neutral decay.
The spectrum is nearly flat in this range, so we may compare 
one half of this number to our theoretical result integrated 
from $q^2=4\,\mbox{GeV}^2$ to $6\,\mbox{GeV}^2$. We obtain 
$(1.2\pm 0.4) \cdot 10^{-7}$ (now including the uncertainty in 
$A_0^{K^*}(0)$), which is only about one half of the central value of
the experimental result. It will be interesting to see whether this 
difference is due to a detection bias, or rather 
points to a problem with the theoretical input. 
 
\begin{figure}[t]
\begin{center}
   \epsfxsize=8.5cm
   \centerline{\epsffile{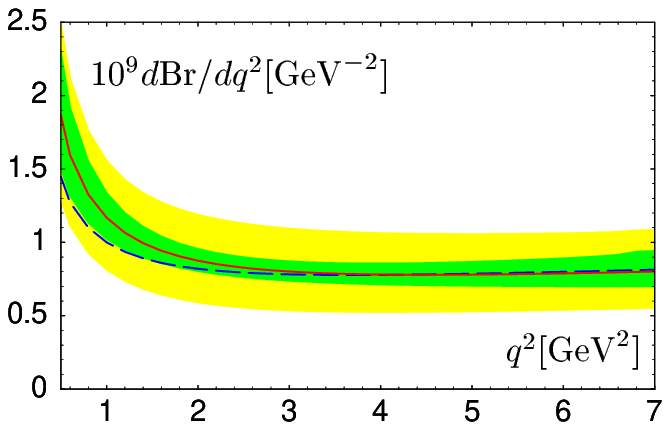}}
\vspace*{-0.7cm}
\end{center}
\caption{CP-averaged differential branching ratio for $B^0\to
\rho^0\ell^+\ell^-$ at NLO as a function of $q^2$ (solid line,
in units of $10^{-9}\,\mbox{GeV}^{-2}$).
The light (yellow) band shows the total theoretical uncertainty.
In the dark (green) band, the uncertainties related to the CKM
parameters and the form factor $A_0^\rho(0)$ are 
excluded. The dashed line shows the LO result.
}
\label{fig:BRrho}
\end{figure}

A new result of the present work is the NNLL (next-to-next-to-leading
logarithmic) prediction for the
$B \to \rho \,\ell^+\ell^-$ decay rate below the charm threshold. 
Integrating the spectrum of the CP-averaged decay as above, we obtain
\begin{eqnarray}
   \int\limits_{1\,{\rm GeV}^2}^{6\,{\rm GeV}^2} \!\!dq^2 \, 
   \frac{d{\rm Br}(B^0\to \rho^0\ell^+\ell^-)}{dq^2}
 &=& 
  \left( \frac{|V_{td}|}{8.25 \cdot 10^{-3}} \,
 \frac{A_0^{\rho}(4\,\mbox{GeV}^2)}{0.50} \right)^{\!2} 
       \, (4.2_{-0.4}^{+0.6}) \cdot 10^{-9},\\[0.2em] 
  \int\limits_{1\,{\rm GeV}^2}^{6\,{\rm GeV}^2} \!\!dq^2 \, 
   \frac{d{\rm Br}(B^+\to \rho^+\ell^+\ell^-)}{dq^2}
 &=& 
  \left( \frac{|V_{td}|}{8.25 \cdot 10^{-3}} \,
 \frac{A_0^{\rho}(4\,\mbox{GeV}^2)}{0.50} \right)^{\!2} 
      \, (9.6_{-1.1}^{+1.5}) \cdot 10^{-9}.
\end{eqnarray}
In Figure~\ref{fig:BRrho} we show the spectrum 
$d{\rm Br}(B^0\to \rho^0\ell^+\ell^-)/dq^2$. The qualitative 
features are similar to $B\to K^*\ell^+\ell^-$, namely the 
NLO correction is very important at small $q^2$, where the 
transverse amplitude is large, but at large $q^2$ the 
spectrum is flat, and the NLO corrections from different sources 
cancel to produce a negligible net effect. 

By the time the $B \to \rho \,\ell^+\ell^-$ decays can be measured, 
the CKM elements, as well as the 
short-distance Wilson coefficients $C_9$ and $C_{10}$ for
$b \to s \ell^+\ell^-$ transitions, will be known with high accuracy. 
The $B \to \rho \,\ell^+\ell^-$ decays can provide additional
insight into the structure of flavour-changing neutral current 
interactions in the context of
new-physics scenarios with non-minimal
flavour violation, where $C_9^d$ and $C_{10}^d$ can differ both,
in magnitude and in phase, from their counterparts in the $b \to s$ sector.
To achieve the necessary theoretical accuracy, it is crucial to
control the hadronic uncertainties. We may expect the form 
factors at small $q^2$ to be known with much better accuracy 
than today, presumably from lattice simulations. In this situation 
the factorization formalism should allow us to obtain 
stringent constraints on  $C_9^d$ and $C_{10}^d$ from the 
neutral decay mode. On the other hand, the theoretical control of 
charged decays remains more challenging because of the leading-power
annihilation effects whose $q^2$ behaviour strongly depends on
the model for the $B$\/-meson wave function.\footnote{These effects
  are suppressed in the branching fraction in the Standard Model, 
  because $\cos\alpha$ is small. This suppression need not hold 
  in extensions of the Standard Model such as discussed here.}

\subsection{Isospin asymmetries}

The isospin asymmetry in $B \to K^*\gamma$ has been discussed
in \cite{Kagan:2001zk}.
Our calculation gives (all rates averaged over the CP-conjugate 
decay)
\beq
  \Delta(K^*\gamma) =
  \frac{
  \Gamma(B^0 \to K^{*0}\gamma) - \Gamma(B^+ \to K^{*+}\gamma)
  }{
  \Gamma(B^0 \to K^{*0}\gamma) + \Gamma(B^+ \to K^{*+}\gamma)}
  = \frac{0.28}{T_1^{K^*}(0)}\, (5.8 ^{+3.3}_{-2.9})\%,
\eeq
which is consistent with the experimental
number $(3.6 \pm 3.8)\%$ deduced from Table~\ref{tabExp}. As already 
discussed, isospin breaking in radiative decays is a power-suppressed 
effect. Although the calculation is believed to capture the dominant 
effect, its theoretical status is less certain than the calculation 
of branching fractions and CP asymmetries. Our result is smaller 
than the result of the original calculation \cite{Kagan:2001zk}, 
because we evaluate the Wilson coefficients at a scale of order 
$m_b$ rather than $(m_b \Lambda_{\rm QCD})^{1/2}$. The motivation 
for this choice is that the four-quark operators factorize 
below the scale $\mu\simeq m_b$, and the gluon exchange between 
all quark lines responsible for the renormalization group running 
of the Wilson coefficients 
is no longer relevant at smaller scales. Our result is larger 
than the one given in \cite{Bosch:2004nd}, partially because in this paper 
the isospin-breaking hard-scattering corrections (denoted by 
``$\alpha_s^1/m_b$'' in Table~\ref{tabCalC7}) are not included, 
but primarily because the larger QCD sum rule value of the $K^*$ 
tensor form factor is used there. 

The extension of the isospin analysis to 
$B \to K^*\ell^+\ell^-$ has been discussed
in \cite{Feldmann:2002iw}. We refrain from giving an updated
discussion here, but only mention that with the different choice
of scale described above, we expect an even smaller isospin asymmetry 
than the estimate in \cite{Feldmann:2002iw}, which implies a larger 
sensitivity to isospin-violating new physics effects.

\begin{figure}[t]
\begin{center}
   \epsfxsize=8.5cm
   \centerline{\epsffile{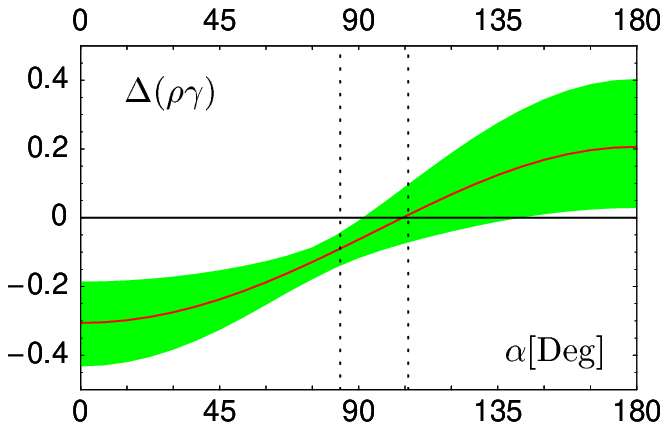}}
\vspace*{-0.7cm}
\end{center}
\caption{\label{fig:Iso} Isospin asymmetry $\Delta(\rho\gamma)$ as a
  function of the CKM angle $\alpha$. 
  The band displays the total theoretical uncertainty which is 
  mainly due to weak annihilation. The vertical dashed lines 
  limit the range of $\alpha$ obtained from the CKM unitarity 
  triangle fit.}
\end{figure}

While the isospin asymmetry in $B \to K^*\gamma$ decays mainly
probes the magnitude of penguin Wilson coefficients in weak
annihilation, the isospin asymmetry for $\rho\gamma$ decays is 
sensitive to CKM parameters through the interference with a large 
tree annihilation amplitude with a different CP phase. 
We obtain for the isospin asymmetry (again an average over the
CP-conjugate decay is understood)
\beq
  \Delta(\rho\gamma) =
  \frac{\Gamma(B^+ \to \rho^+\gamma)}{2\Gamma(
    B^0\to\rho^0\gamma)}-1 =  (-4.6 \, {}^{+5.4}_{-4.2} \big|_{\rm CKM} 
       {}^{+5.8}_{-5.6} \big|_{\rm had})\%.
\label{eq:isodef}
\eeq
The largest hadronic error ($4\%$) comes from the weak
annihilation contribution to which we assign a $50\%$ error. The 
uncertainty labeled ``CKM'' illustrates the sensitivity to 
the CKM parameters. Using the abbreviations (\ref{abbrev1},\ref{abbrev2}),
and neglecting terms quadratic in 
$\delta_+,\epsilon_0\simeq 0.1$ and $\eta$, 
we can approximate the exact result for the isospin asymmetry by 
\begin{eqnarray}
  \Delta(\rho\gamma) &\simeq&    \delta_+
   - 2 \, R_{ut}  \cos\alpha \,
   (\epsilon_+ \cos\theta_+-\epsilon_0 \cos\theta_0)
   + R_{ut}^2 \, \epsilon_+^2
\nonumber\\
   &=& -0.076^{+0.042}_{-0.049} - 
       2 R_{ut} \cos\alpha \,[0.30^{+0.16}_{-0.16}] 
   + R_{ut}^2 \, [0.07^{+0.12}_{-0.06}]
\label{isoapprox}
\end{eqnarray}
with a relative error of less than $10\%$.\footnote{With 
$T_1^\rho(0)=0.21$ instead of 
0.29, the central values of the numerical entries in
(\ref{isoapprox}) change from $(-0.076,0.30,0.07)$ to 
$(-0.104,0.42,0.18)$.} For $\alpha$ near
$90^\circ$ the isospin asymmetry is predicted to be small, and 
the numerical result is dominated by the isospin asymmetry 
in the top quark sector, $\delta_+$. Our result (\ref{eq:isodef}) 
therefore differs from~\cite{Ali:2004hn}, where 
the term $\delta_+$ is neglected. Eq.~(\ref{isoapprox}) also 
shows that there is a large sensitivity to $\cos\alpha$, 
although with a large uncertainty related to the weak annihilation
amplitude. This dependence is shown in Figure~\ref{fig:Iso}. 

\begin{figure}[t]
\begin{center}
   \epsfxsize=8.5cm
   \centerline{\epsffile{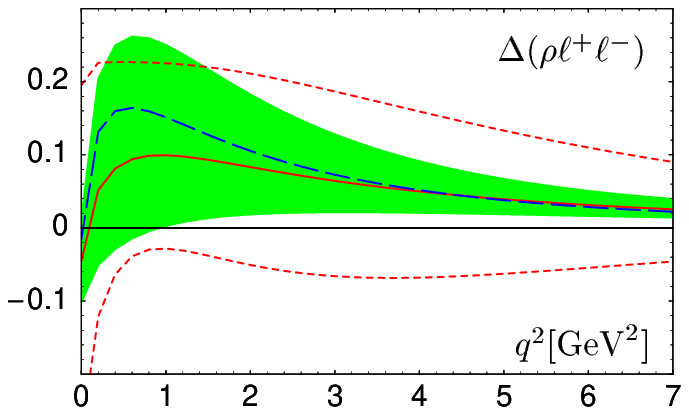}}
\vspace*{-0.7cm}
\end{center}
\caption{Isospin asymmetry $\Delta(\rho\,\ell^+\ell^-)$ as a
  function of $q^2$. The solid (long-dashed) line shows the 
  next-to-leading (leading) order result for $\alpha=94^\circ$. 
  The band represents 
  the hadronic uncertainty. The two dashed lines give  
  $\Delta(\rho\,\ell^+\ell^-)$ for $\alpha=24^\circ$ (lower curve) and 
  $\alpha=164^\circ$ (upper curve).}
\label{fig:isoq2}
\end{figure}

The isospin asymmetry in the differential decay spectrum
for $B \to \rho\,\ell^+\ell^-$ is defined in analogy to
(\ref{eq:isodef}). In the limit $q^2\to 0$ its numerical 
value approaches (\ref{isoapprox}). For invariant lepton-pair masses 
well above the photon pole at $q^2=0$ the decay spectrum
is dominated by the longitudinal rate, and therefore the
isospin asymmetry mainly comes from the amplitudes 
${\cal C}_{9,\parallel}^{(i)}$ with $i=t,u$. It can be 
seen from Table~\ref{tabCalC9} that the dominant effect comes from 
the leading-power weak annihilation contribution~\cite{Beneke:2001at}
to the charged decay mode
$B^+ \to \rho^+\ell^+\ell^-$ which, as already explained,
has a large uncertainty. When $\cos\alpha$ is not near zero, 
the asymmetry is generated mainly by ${\cal C}_{9,\parallel}^{(u)}$ 
and the sign of the asymmetry is determined by $-\cos\alpha$ 
just as for $B\to\rho\gamma$. On the other hand, for small 
$\cos\alpha$ (expected in the Standard Model), the asymmetry 
is small, comes mainly from the top quark sector, and 
its sign is opposite to (\ref{eq:isodef}). In any case, the 
effect is diluted compared to $B\to\rho\gamma$ because of 
the isospin-symmetric contribution of the Wilson coefficient 
$C_{10}$ to the longitudinal amplitude. This situation is
illustrated in Figure~\ref{fig:isoq2}, where we show the 
predicted asymmetry for $\alpha=24^\circ$, $164^\circ$ 
and $94^\circ$ (default). For the last value of $\alpha$ 
the theoretical uncertainty is also shown. The maximum of the 
isospin asymmetry around $q^2\simeq 1\,\mbox{GeV}^2$ can be 
explained by the observation that the longitudinal amplitude 
for the charged decay 
becomes dominated by the leading-power weak annihilation contribution
for $q^2\to 0$, since $\lambda_{B,-}(q^2)^{-1}$ increases 
logarithmically. Hence the isospin asymmetry increases (coming from 
larger $q^2$) until the transverse decay amplitude wins over 
and the isospin asymmetry turns to the negative value  
(\ref{isoapprox}) for $q^2=0$. A reliable prediction is currently
possible only in the larger $q^2$ region shown in the Figure. 

As mentioned above the $B\to\rho\,\ell^+\ell^-$ decays are 
mainly of interest in the context of scenarios where physics beyond the
standard model modifies the electroweak penguin coefficients 
$C_9$, $C_{10}$ in $b\to d$ transitions. If $\theta$ denotes the extra
CP-phase of $C_9$, the isospin asymmetry depends, roughly speaking, on 
$\alpha_{\rm eff}=\alpha-\theta$. Since $\alpha$ will be known, 
the sensitivity of the isospin asymmetry to $\alpha$ shown in 
Figure~\ref{fig:isoq2} translates into a 
sensitivity to $\theta$. The theoretical uncertainty of the 
Standard Model reference value due to weak annihilation implies 
that only a significant new phase could be unambiguously detected. 
In this case a significant modification of the absolute value 
of $C_9$ and $C_{10}$ is also 
probable and might be directly observable in the 
lepton invariant mass spectra.

\subsection{Direct CP asymmetries}
\label{sec:3.4}

\begin{figure}[t]
\begin{center}
   \epsfxsize=8.5cm
   \centerline{\epsffile{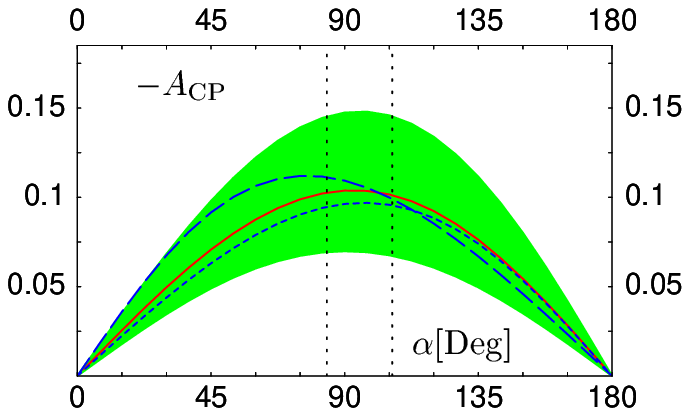}}
\vspace*{-0.7cm}
\end{center}
\caption{\label{fig:CPgam} 
  Direct CP asymmetry in $B\to\rho^0\gamma$ (solid), 
  $B^+\to\rho^+\gamma$ (long-dashed) and
  $B\to\omega\gamma$ (short-dashed) decay as a
  function of the CKM angle $\alpha$. 
  The band shows the theoretical uncertainty for the case of 
  $B\to\rho^0\gamma$. Note that we display {\em minus} the CP 
  asymmetry.}
\end{figure}

The direct CP asymmetries for $B \to \rho\gamma$ decays 
are given by 
\beq
  A_{\rm CP}(\rho^{0}\gamma) =  \frac{
  \Gamma[\bar B^0 \to \rho^0\gamma] - \Gamma[B^0 \to \rho^0\gamma]
  }{
  \Gamma[\bar B^0 \to \rho^0\gamma] + \Gamma[B^0 \to \rho^0\gamma]}
  = \frac{2 R_{ut} \epsilon_0  \sin\theta_0 \sin\alpha}
  {1-2 R_{ut} \epsilon_0 \cos\theta_0 \cos\alpha+R_{ut}^2 \epsilon_0^2}, 
\eeq
and an analogous equation for the charged $B$ decay with 
$(\epsilon_0,\theta_0)\to(\epsilon_+,\theta_+)$. The values 
of $\epsilon_i\cos\theta_i$ and $\epsilon_i^2$ have already been 
given in (\ref{br_neutral},\ref{br_charged}). Here we also need    
\begin{equation}
\epsilon_0 \sin\theta_0 = -0.11^{+0.03}_{-0.04}, 
\qquad
\epsilon_+ \sin\theta_+ = -0.12^{+0.03}_{-0.04}.
\label{imeps}
\end{equation}
The largest theoretical uncertainties are from the residual
renormalization scale dependence ($\pm 0.03$) and the charm quark 
mass ($\pm 0.02$). This reflects the fact that a next-to-leading order
calculation of the branching fractions results in leading-order
predictions of direct CP asymmetries, which are therefore more
sensitive to unknown higher-order and power corrections. 
It is worth noting that the product 
$\epsilon_+ \sin\theta_+$ is less dependent on the weak annihilation 
contribution than the individual factors $\epsilon_+$ and $\sin \theta_+$, 
since the annihilation amplitude has no strong phase relative to the
leading electromagnetic penguin amplitude. It follows from 
(\ref{imeps}) that, within uncertainties,
the CP asymmetries in neutral and charged $B \to\rho\gamma$
decays are of similar size
\begin{eqnarray}
  A_{\rm CP}(\rho^{0}\gamma) &=& 
   (-10.4 \, {}^{+1.6}_{-2.4} \big|_{\rm CKM} 
       {}^{+3.0}_{-3.6} \big|_{\rm had})\,\%,
\nonumber\\
  A_{\rm CP}(\rho^{+}\gamma) &=&    
   (-10.7 \, {}^{+1.5}_{-2.0} \big|_{\rm CKM} 
       {}^{+2.6}_{-3.7} \big|_{\rm had})\,\% . 
\label{directacp}
\end{eqnarray}
The dominant dependence of the CP asymmetries on the CKM parameters 
is through $R_{ut} \sin\alpha$. The corresponding constraint in the 
$(\bar\rho,\bar\eta)$ plane is discussed in Section~\ref{sec:3.5}. 
In Figure~\ref{fig:CPgam} we 
show the dependence of the direct CP asymmetries on 
the CKM angle $\alpha$. The asymmetries for $B\to\rho^0\gamma$ and 
$B\to\omega\gamma$ are indistinguishable within uncertainties. Our
result is in agreement with \cite{Bosch:2001gv,Ali:2004hn} within 
theoretical uncertainties, though \cite{Ali:2004hn} displays a
slightly larger difference between the neutral and charged CP 
asymmetries.

\begin{figure}[t]
\begin{center}
   \epsfxsize=8.5cm
   \centerline{\epsffile{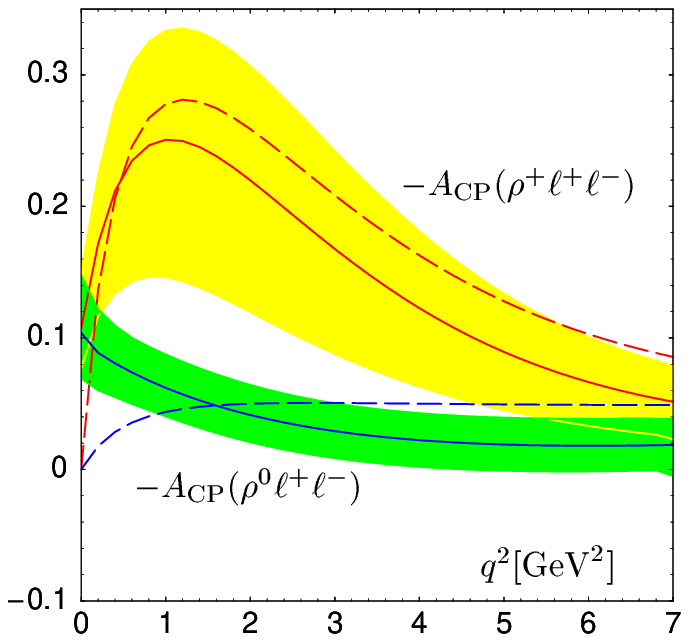}}
\vspace*{-0.7cm}
\end{center}
\caption{Direct CP asymmetries in $B^0 \to \rho^0\ell^+\ell^-$ 
(lower set of curves) and $B^+ \to \rho^+\ell^+\ell^-$ as a function
of $q^2$. The solid (dashed) curves show the next-to-leading order
(leading order) result. The widths of the bands represent the 
hadronic uncertainty. Note that we display {\em minus} the CP asymmetry.}
\label{fig:CPneutral}
\end{figure}

The direct CP asymmetry arises in $B \to \rho\,\ell^+\ell^-$ decays
from the interference between the ${\cal C}_{9,a}^{(u)}(q^2)$ and
${\cal C}_{9,a}^{(t)}(q^2)$ amplitudes with different strong
phases. From Table~\ref{tabCalC9} we deduce that the largest contributions
to strong phases come from the one-loop function $Y(q^2)$,
from the coefficient function $C^{(1)}$, and, in the 
case of the  charged decay, from
annihilation topologies which involve the 
$B \to \gamma^*$ form factors $\xi_{\perp,\parallel}^{B\gamma^*}(q^2)$
(see appendix). The two-loop virtual correction $C^{(1)}$ calculated in
\cite{Seidel:2004jh,Asatrian:2003vq} plays an essential role here,
since it cancels a large part of 
the imaginary part of $Y^{(u)}(q^2)$ in ${\cal C}_{9,\perp}^{(u)}$ and
${\cal C}_{9,\parallel}^{(u)}$. 
For the charged decay  $B^\pm\to\rho^\pm \ell^+\ell^-$ the residual 
relative strong phase is then
dominated by the rather uncertain annihilation effects. 
For the neutral decay $B^0\to\rho^0 \ell^+\ell^-$, on the contrary, 
the strong phases are expected to be rather small, leading to a small 
CP asymmetry within the Standard Model.
Our numerical prediction for the two decay modes 
is shown in Figure~\ref{fig:CPneutral}. The increase of 
$-A_{\rm CP}(\rho^+\ell^+\ell^-)$ near $q^2\simeq 1\,\mbox{GeV}^2$ 
occurs for the same reason as discussed for the isospin asymmetry 
and is correspondingly uncertain. 
In extensions of the Standard 
Model $\alpha$ is again replaced by $\alpha_{\rm eff}$, but given the 
theoretical errors and given that the CP asymmetry is expected to be
nearly maximal in the Standard Model, it will be difficult to
disentangle a new phase unless the CP asymmetry in the charged 
decay mode is strongly suppressed. 

\subsection{CKM constraints}
\label{sec:3.5}

As seen in previous sections, measurements of the branching ratios, 
isospin- and CP-asymmetries
in $B \to \rho\gamma$ decays are sensitive to the CKM elements and
thus provide interesting constraints on the parameters $\bar \rho$
and $\bar \eta$, which define the apex of the unitarity triangle. 
In this section we summarize these constraints. 

\begin{figure}[t]
\begin{center}
\hskip-3.5cm
\psfig{file=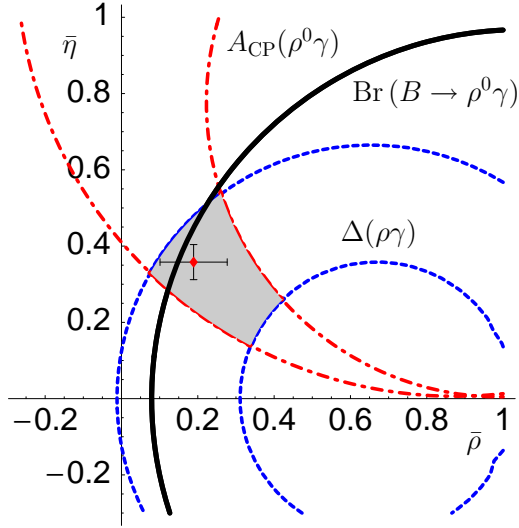, width=0.6 \textwidth}
\end{center}
\vskip-0.2cm
\caption{Constraints on the unitarity triangle from the ratio of 
the CP-averaged $B^0\to\rho^0\gamma$ and $B^0\to K^{*0}\gamma$
branching fractions (solid black), the isospin asymmetry in 
$B\to\rho\gamma$ (dashed) and the direct CP asymmetry in
$B^0\to\rho^0\gamma$ (dash-dotted). The area to the left of 
the black line is excluded by the experimental limit on 
$B^0\to\rho^0\gamma$. 
See text for explanations.}
\label{fig:ckm}
\end{figure}

At present, there exists only an upper experimental
limit for the $B  \to \rho\gamma$ branching fractions.
Interestingly, our theoretical result (\ref{Brho0BR}) already
saturates the experimental bound, and therefore detection of 
this decay is expected in the very near future. The existing upper limit 
on $B \to \rho^0\gamma$ translates into a useful bound on $|V_{td}|$, 
which is complementary 
to the bound from the non-observation of $B_s\bar B_s$-mixing. 
The key point is that the curly bracket in (\ref{br_neutral}) 
is constrained to be close to 1. Adding the dominant 
parameter dependencies linearly and doubling the error 
on the input values of $(\bar\rho,\bar\eta)$ we 
find 
\begin{equation} 
0.94 <  1 - 2 R_{ut} \, \epsilon_0 \cos\alpha \, \cos\theta_0
   + R_{ut}^2 \, \epsilon_0^2  <1.05.
\label{curlybracket}
\end{equation}
Using the data from Table~\ref{tabExp} this translates into 
the bound 
\begin{equation}
 \left| \frac{V_{td}}{V_{ts}} \right| = 
\frac{1\pm 0.03}{F} \left(\frac{2 {\rm Br}(B^0 \to \rho^0\gamma)}
       {{\rm Br}(B^0\to  K^{*0}\gamma)}\right)^{1/2} 
 < 0.21,
\label{vtdbound}
\end{equation}
where we used the conservative range $F>0.7$, which gives the weakest 
bound. A similar bound has been obtained 
in \cite{Bosch:2004nd}.\footnote{ 
In~\cite{Ali:2004hn} an average over $\rho^0,\rho^\pm,\omega$ 
is performed, which reduces the significance of the constraint due 
to the larger theoretical uncertainty in 
$B\to\rho^\pm\gamma$ and the weaker experimental upper limit 
on the $\rho^\pm$ and $\omega$ final states.}
This limit already cuts into the range 
$ \left|V_{td}/V_{ts}\right| = 0.204^{+0.029}_{-0.046}$ 
(CL=0.05) obtained from the standard 
fit~\cite{Charles:2004jd}. This is displayed in Figure~\ref{fig:ckm}, 
where the area to the left of the solid black curve is excluded 
by (\ref{vtdbound}), and the standard values of $(\bar\rho,\bar\eta)$ 
are shown as a point together with their errors as 
in Table~\ref{tabInput}.\footnote{The solid line becomes 
slightly inaccurate far away from the standard range, since 
the dependence of (\ref{curlybracket}) on $(\bar\rho,\bar\eta)$ 
is neglected.}

No data currently exists for the isospin and direct CP asymmetries. 
To illustrate the possible constraints we assume that they 
are observed with a value that corresponds to our theoretical 
expectations (\ref{isoapprox},\ref{directacp}) without experimental 
error. The theoretical uncertainty from the input parameters 
(excluding CKM parameters)  
then translates into the dashed (isospin asymmetry) and 
dash-dotted (direct CP asymmetry in $B\to\rho^0\gamma$) bands 
in the Figure. The constraint from the CP asymmetry in the charged 
decay $B^\pm\to\rho^\pm\gamma$ is very similar to the neutral one 
in the vicinity of the standard $(\bar\rho,\bar\eta)$ range 
and not shown. The shape of these constraints can be understood 
from the relations 
\begin{equation}
  R_{ut} \sin\alpha =  \frac{\bar\eta}{(1-\bar\rho)^2 + \bar \eta^2}, 
\qquad 
  R_{ut}\cos\alpha =  1 - \frac{1-\bar\rho}
  {(1-\bar\rho)^2 + \bar \eta^2}.
\end{equation}
They imply to good approximation that the direct CP asymmetry 
requires $(\bar\rho,\bar\eta)$ to lie on a circle of 
radius $1/(2 a)$ with center $(1,1/(2 a))$ where  
\begin{equation}
a = \frac{A_{\rm CP}(\rho^{0}\gamma)}{2 \epsilon_0 \sin\theta_0}, 
\end{equation}
which always passes through $(1,0)$. 
Similarly, the isospin asymmetry constrains $(\bar\rho,\bar\eta)$ to
lie on a circle whose center is always at $\bar\eta=0$ and whose
radius is determined from (\ref{isoapprox}). We see from 
Figure~\ref{fig:ckm} that the two asymmetry constraints intersect 
nearly orthogonally, the intersection region being shaded 
in grey. This area is further constrained by the limit on the 
$B\to\rho\gamma$ branching fraction. The three observables together 
(and a similar constraint from the direct asymmetry in $B^\pm\to
\rho^\pm \gamma$) demonstrate that the $B\to\rho\gamma$ decays 
alone provide valuable independent information on CKM parameters. 
Furthermore, if inconsistencies with the standard CKM fit appeared, 
this would point towards anomalous effects in $b\to d\gamma$
transitions.

\subsection{Forward-backward asymmetries}
\label{sec:3.6}

The forward-backward asymmetry in $B\to V\ell^+\ell^-$ decays is
defined by 
\begin{eqnarray}
 A_{\rm FB}(q^2) &\equiv &
 \frac{1}{d\Gamma/dq^2} \left(\,
\int_0^1 d(\cos\theta) \,\frac{d^2\Gamma}{dq^2 
d\cos\theta} - \int_{-1}^0 d(\cos\theta) \,\frac{d^2\Gamma}{dq^2 
d\cos\theta} \right) 
 \nonumber\\[0.2cm]
 &\propto& \mbox{Re}\left({\cal C}_{9,\,\perp}^{(t)}(q^2)
  -R_{ut}\, e^{\pm i\alpha}\,{\cal C}_{9,\,\perp}^{(u)}(q^2)
  \right),
\end{eqnarray}
where the second line follows from (\ref{gammadist}). Note that 
we have {\em not} performed an average over CP-conjugate decays 
for the forward-backward asymmetry. The exponential reads 
$e^{-i\alpha}$ for $B$ decay, and $e^{i\alpha}$ for $\bar B$ decay. 

The next-to-leading order prediction of the 
forward-backward asymmetry for the $B \to K^* \ell^+\ell^-$ decay
has been discussed in detail in our previous 
paper \cite{Beneke:2001at}. For the $b\to s$ transitions 
the term ${\cal C}_{9,\,\perp}^{(u)}(q^2)$ is negligible, because 
the corresponding $R_{ut}$ 
is very small. Hence there is no difference between $B$ and $\bar B$
decay, and the asymmetry zero is determined by the 
zero of the real part of ${\cal C}_{9,\,\perp}^{(t)}(q^2)$. 
In \cite{Beneke:2001at} we found 
that the next-to-leading order correction shifts the zero by 
$30\%$, but once this correction is included, a precise 
measurement of the location of the zero translates into 
a determination of the Wilson coefficient $C_9$ with an accuracy 
of about $10\%$. Our updated result for the position of the 
forward-backward asymmetry zero reads 
\beq
  q_0^2[K^{*0}] = 4.36^{+0.33}_{-0.31} \,\mbox{GeV}^2,\qquad
  q_0^2[K^{*+}] = 4.15^{+0.27}_{-0.27} \, \mbox{GeV}^2. \qquad
\eeq
The small difference compared to~\cite{Beneke:2001at} is 
due to the different treatment 
of form factors and the inclusion of isospin breaking power
corrections  in the present analysis.

\begin{figure}[t]
\begin{center}
\psfig{file=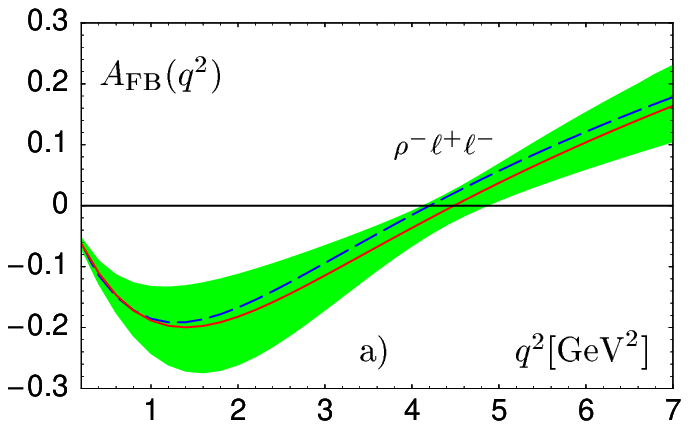}
\psfig{file=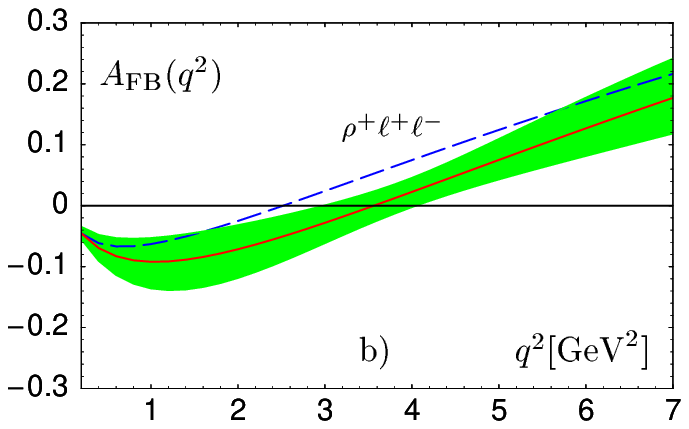}
\vskip0.5cm
\psfig{file=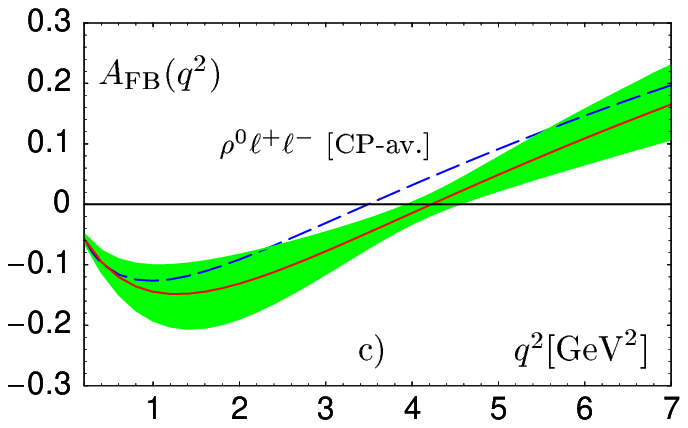}
\end{center}
\vskip-0.2cm
\caption{The forward-backward asymmetry in 
a) $B^+ \to \rho^+\ell^+\ell^-$, b) $B^- \to \rho^-\ell^+\ell^-$,  
and c) the CP-averaged $B \to \rho^0\ell^+\ell^-$ decay. The solid 
(dashed) line shows the next-to-leading (leading) order result. 
The band represents the theoretical error due to hadronic 
uncertainties.}
\label{fig:FBrho}
\end{figure}
In case of $B \to \rho\,\ell^+\ell^-$ decays there exists an important 
new contribution from  ${\cal C}_{9,\,\perp}^{(u)}(q^2)$. 
As a consequence, the decays of $B$ or $\bar B$, neutral or charged 
$B$ mesons to $\rho\,\ell^+\ell^-$ may show significantly different 
forward-backward asymmetries. When $\alpha$ is near $90^\circ$ as 
expected in the Standard Model, we may approximate 
$e^{i\alpha} \simeq i \sin\alpha$, and therefore the additional 
contribution to the forward-backward
asymmetry involves approximately 
\begin{equation}
 R_{ut}\sin\alpha\,{\rm Im}\,\Big({\cal C}_{9\perp}^{(u)}(q^2)\Big),
\end{equation}
i.e.~the absorptive part of the amplitude from the up-sector. 
We see from Table~\ref{tabCalC9} that large contributions to the
absorptive part arise from the virtual corrections to the 
$b\to d\ell^+\ell^-$ transition, $Y^{(u)}(q^2)$ and $C^{(1)}$, and that, 
in particular, the two-loop correction $C^{(1)}$ calculated in
\cite{Seidel:2004jh,Asatrian:2003vq} is as large as the one-loop 
term $Y^{(u)}(q^2)$. The other large contribution comes from weak 
annihilation and exists only for charged $B$ decays. 
In Figure~\ref{fig:FBrho} we show the expected forward-backward
asymmetries for $B^\pm\to\rho^\pm\,\ell^+\ell^-$ and the asymmetry for 
the {\em CP-averaged} $B\to\rho^0\,\ell^+\ell^-$ decay
rate.\footnote{The case of $\omega$ is not 
significantly different from $\rho^0$.}  
The locations of the asymmetry zeros are determined as 
\begin{equation}
\begin{array}{cccc}
& B^0\to \rho^0\,\ell^+\ell^- &\phantom{b}& 
\bar B^0\to \rho^0\,\ell^+\ell^-\\[0.2cm]   
& 4.34 \, {}^{+0.02}_{-0.02} \big|_{\rm CKM} 
       {}^{+0.45}_{-0.39} \big|_{\rm had} && 
  4.11 \, {}^{+0.02}_{-0.02} \big|_{\rm CKM} 
       {}^{+0.26}_{-0.22} \big|_{\rm had} \\[0.1cm]
q_0^2[\mbox{GeV}^2]\hskip0.5cm & & \\[-0.1cm]
 & B^+\to \rho^+\,\ell^+\ell^- && B^-\to \rho^-\,\ell^+\ell^- \\[0.2cm]

& 3.56 \, {}^{+0.09}_{-0.12} \big|_{\rm CKM} 
       {}^{+0.48}_{-0.57} \big|_{\rm had} && 
  4.48\, {}^{+0.07}_{-0.06} \big|_{\rm CKM} 
       {}^{+0.36}_{-0.34} \big|_{\rm had} 
\end{array}
\label{asym0s}
\end{equation}
As in the case of $B\to K^*\ell^+\ell^-$, the location of the asymmetry zero 
is a measure of $C_9$ (now in the $b\to d$ sector), but in addition 
information about the phase can be obtained through the interference 
with the tree-dominated amplitude in the up-sector. 

\section{Conclusions}
\label{sec:conclude}

Our analysis provides Standard Model expectations for 
the exclusive, radiative and electroweak penguin decays
$B \to V\gamma$ and $B \to V \ell^+\ell^-$ ($V
=K^*,\rho,\omega$), extending our previous work~\cite{Beneke:2001at} 
to $b \to d\gamma$ and $b \to d\ell^+\ell^-$ transitions. The 
theoretical framework is complete at next-to-leading order 
in the strong coupling (except for weak annihilation which 
is included only in leading order), and leading 
power in the heavy-quark expansion. We have also included
$1/m_b$ power corrections, mainly from weak annihilation, 
which are important for the isospin asymmetries.

The observables related to $B \to \rho\gamma$ decays
can provide interesting constraints on the 
CKM triangle as was also recently discussed in 
\cite{Ali:2004hn,Bosch:2004nd}. The accuracy of these 
constraints is limited by hadronic uncertainties, mainly 
from the  hadronic form factors describing the $B \to \rho$ 
transition (for branching fractions), $B \to \gamma$ transitions/weak  
annihilation (for the isospin asymmetry), 
and from the scale and charm quark mass 
uncertainty (for direct CP asymmetries). 
Nevertheless, the simultaneous measurement of
branching fractions, isospin- and CP asymmetries can be used for
an independent determination of the apex $(\bar \rho, \bar \eta)$
of the CKM triangle, which is complementary to the standard fit. 
In particular, if $B^0\to\rho^0\gamma$ is not observed with 
a branching fraction near the current upper limit 
$4\cdot 10^{-6}$, a tension with the standard fit arises. 

We compared our calculations for $B \to K^*\ell^+\ell^-$ 
with the first experimental results on these decays. The central value of 
the Belle data on the partially integrated lepton invariant mass  
spectrum is about a factor of 2 larger than the theoretical 
prediction. Although the discrepancy is not significant within current
uncertainties, it will be interesting to see how it develops. 
If the problem is theoretical, this may result in the curious 
situation that the comparison with data seems to favour a larger 
form factor $A_0$, but a smaller tensor form factor $T_1$. 
The present experimental accuracy for the measurement of
the forward-backward lepton asymmetry in
this decay mode is not yet competitive with the theoretical one.
In the future, however, we expect significant constraints on the 
Wilson coefficient $C_9$ in $b \to s \ell^+\ell^-$ transitions
as already discussed in \cite{Beneke:2001at}. 

We performed the first next-to-leading order calculation of 
exclusive $b\to d\ell^+\ell^-$ transitions. In the long run when
detailed experimental information on these transitions will become 
available, the CKM matrix can be assumed to be known. Hence 
the various observables related to $ B \to \rho\,\ell^+\ell^-$
decays provide information on the modulus and
phase of the Wilson coefficient $C_9$ in $b \to d\ell^+\ell^-$
transitions or related coefficients in extended flavour models, 
which is of interest in the context of new-physics
scenarios with non-minimal flavour violation. The theoretical
uncertainty related to hadronic input is currently sizeable, 
but we may look forward to improved determinations of $B\to\rho$ 
form factors, and perhaps, the mechanism of weak annihilation, 
on the time scales of interest. These improvements will add $B \to
\rho\,\ell^+\ell^-$ to the list of rare processes that play an 
essential role in the program to uncover the origin of 
flavour violation. 

\vskip0.4cm\noindent
{\bf Acknowledgements}

\vskip0.2cm\noindent 
We would like to thank Patrick Koppenburg for helpful discussions, 
and Stefan Bosch for reading the manuscript.  
This work was supported by the DFG 
Sonderforschungsbe\-reich/Transregio 9 ``Computergest\"utzte Theoretische 
Teilchenphysik''. M.B. would like to thank the INT, Seattle for its 
generous hospitality while part of the work was being done. 
D.S. acknowledges support of the 
DFG Graduiertenkolleg ``Elementarteilchenphysik an der TeV-Skala'' 
and would like to thank the CERN Theory group for 
hospitality. 

\begin{appendix}

\section{The amplitudes ${\cal T}_{a}^{(u,t)}$}
\label{app:a}

In this appendix we present the generalization of the amplitudes 
${\cal T}_{a}^{(t)}$ ($a=\perp,\parallel$) given in~\cite{Beneke:2001at} for 
$B\to K^*\ell^+\ell^-$ to the case of $\rho$ or 
$\omega$ in the final state, and the new amplitudes ${\cal T}_{a}^{(u)}$ 
as defined in (\ref{Matrixel}). In~\cite{Beneke:2001at} we used 
a set of Wilson coefficients denoted by $\bar C_i$. The exact 
relation between the $\bar C_i$ and the $C_i$ given in 
Table~\ref{tabWilson} can be found in the appendix of~\cite{Beneke:2001at}.

The coefficient functions  $C_a^{(i)}$ and
$T^{(i)}_{a,\,\pm}$ 
defined in~(\ref{calT}) have the expansions 
\begin{eqnarray}
C_a^{(i)} &=& C_a^{(0,i)}+\frac{\alpha_s C_F}{4\pi} \,C_a^{(1,i)}
+ \ldots,
\label{c11}
\\[0.2em]
T^{(i)}_{a,\,\pm}(u,\omega)&=&
T^{(0,i)}_{a,\,\pm}(u,\omega)+
\frac{\alpha_s C_F}{4\pi} \,T^{(1,i)}_{a,\,\pm}(u,\omega)
+ \ldots.
\label{t11}
\end{eqnarray}
In the following we give the expressions for the coefficients 
$C_a^{(i)}$ and $T^{(i)}_{a,\,\pm}$. The strong coupling is evaluated
at the scale $\mu\simeq m_b$ in (\ref{c11}) and at 
$\mu_f = (0.5\,\mbox{GeV} \mu)^{1/2}\simeq (m_b\Lambda_{\rm
  QCD})^{1/2}$ in (\ref{t11}), which corresponds to the typical 
virtualities in the two terms. In contrast to our earlier analysis, 
we always evaluate 
the Wilson coefficients at $\mu\simeq m_b$, since the running of the
four-quark operators ends at this scale.

\subsection{Form-factor term}

The coefficients $C_a^{(0,t)}$ follow from Eqs.~(12,14) 
in~\cite{Beneke:2001at}. The corresponding expressions for 
$C_a^{(0,u)}$ are obtained by the replacements $C_7^{\rm eff} \to 0$ 
and 
\begin{equation}
Y(s)\to Y^{(u)}(s) \equiv \left( \frac43 C_1+C_2\right)
  \left[h(s,m_c) - h(s,0)\right],
\end{equation}
where $h(s,m_q)$ is given in Eq.~(11) of~\cite{Beneke:2001at}.

The first-order corrections $C_a^{(1,i)}$ are divided into a 
``factorizable'' and a ``non-factor\-izable'' term according to 
$C_a^{(1,i)} = C_a^{(f,i)} + C_a^{(nf,i)}$. The factorizable
correction reads 
\begin{eqnarray}
\label{Cperpfac}
  C_\perp^{(f,t)} &=& C_7^{\rm eff} 
  \left(\ln \frac{m_b^2}{\mu^2} - L + \Delta M\right),
\\[0.2em]
\label{Cpartfac}
  C_\parallel^{(f,t)} &=& - C_7^{\rm eff} \left(
   \ln \frac{m_b^2}{\mu^2} + 2 L  + \Delta M\right)
\end{eqnarray}
with $L$ defined in Eq.~(36) of~\cite{Beneke:2001at}. $\Delta M$ 
depends on the mass renormalization convention for the overall factor 
$m_b$ in (\ref{Matrixel}), such that $\Delta M=0$ in the 
$\overline{\rm MS}$ scheme, $\Delta M=3\ln (m_b^2/\mu^2)-4
(1-\mu_f/m_b)$ in the PS scheme (our choice) and 
$\Delta M=3\ln (m_b^2/\mu^2)-4$ in the pole mass scheme. Note 
that the expression for $C_\parallel^{(f,t)}$ differs from 
Eq.~(35) of~\cite{Beneke:2001at} due to the different convention 
for the longitudinal ``soft'' form factor (\ref{xiperp}). 
Furthermore, when ${\cal T}^{(t)}_\perp$ is defined 
with $T_1$ rather than $\xi_\perp$, the corresponding 
$C_\perp^{(f,t)\prime}$ is 
given by (\ref{Cperpfac}) with 
the term $-L$ omitted. The factorizable corrections from 
$H_{\rm eff}^{(u)}$ follow 
again from  $C_7^{\rm eff} \to 0$, which leads to 
$C_\perp^{(f,u)} = C_\parallel^{(f,u)} = 0$.

The non-factorizable corrections $C_a^{(nf,t)}$ are given 
in Eqs.~(37,38) of~\cite{Beneke:2001at}, and make use of the 
result from~\cite{Asatryan:2001zw}. The corresponding expressions for 
$C_a^{(nf,u)}$ are obtained by the replacements 
$F_8^{(7,9)}\to 0$ and $F_{1,2}^{(7,9)}\to 
F_{1,2}^{(7,9)} + F_{1,2,u}^{(7,9)}$ 
with $F_{1,2,u}^{(7,9)}$ given in~\cite{Seidel:2004jh}.  

\subsection{Spectator scattering} 

The longitudinal amplitude receives a leading-order contribution
from a weak annihilation topology, where the photon couples to 
the spectator quark in the $B$ meson~\cite{Beneke:2001at}. 
It is given by ($e_q$ denotes the charge of the spectator quark)
\begin{eqnarray}
T_{\parallel,-}^{(0,t)}(u,\omega) 
= - e_q \, \frac{M_B \omega}{M_B \omega - q^2 - i\epsilon} \,
   \frac{4M_B}{m_b} \, C_q^{34}, 
\label{annlong} 
\end{eqnarray}
where 
\begin{equation}
C_q^{34} \equiv 
     C_3 + \frac43 ( C_4 + 12 C_5 + 16 C_6 )
 + 6 \, [1 + (-1)^I]  \,  \delta_{qd} \,
     ( C_3 + 10 C_5 )
\end{equation}
with $I=0,1$ for $\omega$ and $\rho$ mesons, respectively. 
The corresponding expression for $T_{\parallel,-}^{(0,u)}$ is 
obtained by the replacement $C_q^{34}\to - C_q^{12}$ with 
\begin{equation}
C_q^{12} \equiv 
   3 \, \delta_{qu} \,  C_2
 + (-1)^{I} \, \delta_{qd} \, \left( \frac43 \, C_1 + C_2 \right). 
\end{equation}
The terms proportional to $\delta_{qd}$ in the previous two equations 
do not appear in $B \to K^*\ell^+\ell^-$ transitions. 
(Analogous terms
with $\delta_{qd} \to \delta_{qs}$ would appear in
$B_s \to \phi \ell^+\ell^-$.) There is no leading-order contribution 
from spectator scattering to $T_{\parallel,+}^{(0,i)}$ 
and the transverse amplitudes. 

The first-order corrections are again divided into a ``factorizable''
and a ``non-factor\-izable'' term, $T_{a,\pm}^{(1,i)} = 
T_{a,\pm}^{(f,i)} + T_{a,\pm}^{(nf,i)}$. There is also a first-order 
correction to the annihilation mechanism discussed above, which is not yet
known. $T_{\perp,\pm}^{(f,t)}$ and $ T_{\parallel,-}^{(f,t)}$ 
are given in Eqs.~(20,22) of~\cite{Beneke:2001at}, while now 
\begin{equation}
\label{Tpartfac}
  T_{\parallel,+}^{(f,t)} =
  C_7^{\rm eff} \, \frac{4 M_B}{\bar u E},
\end{equation}
because of the different convention pertaining to 
the longitudinal ``soft'' form 
factor.  The corresponding terms from 
$H_{\rm eff}^{(u)}$ all vanish, $T_{a,\pm}^{(f,u)} = 0$. 
Furthermore, when ${\cal T}^{(t)}_\perp$ is defined 
with $T_1$ rather than $\xi_\perp$, the corresponding 
$T_{\perp,+}^{(f,t)\prime}$ is zero.

The non-factorizable amplitudes $T_{a,\pm}^{(nf,t)}$ are given in 
Eqs.~(23-26) of~\cite{Beneke:2001at}. The new amplitudes read 
\begin{eqnarray}
  T_{\perp,+}^{(nf,u)}(u,\omega) &=&
   e_u \, \frac{M_B}{2m_b} 
    \,  \left(C_2 - \frac16 C_1\right)
    \left( t_\perp(u,m_c)-t_\perp(u,0) \right),
\nonumber \\[0.2em]
  T_{\parallel,+}^{(nf,u)}(u,\omega) &=&
  e_u \, \frac{M_B}{m_b} \, \left(C_2 -\frac16 C_1 \right)
  \left(t_\parallel(u,m_c) - t_\parallel(u,0) \right),
\nonumber \\[0.2em]
  T_{\parallel,-}^{(nf,u)}(u,\omega) &=&
 e_q \, \frac{M_B \omega}{M_B \omega - q^2 - i\epsilon}
  \, \frac{6 M_B}{m_b} \CR
  &&\times\left(C_2 -\frac16 C_1 \right)
  \left( h(\bar u M_B^2 + u q^2,m_c) -  
         h(\bar u M_B^2 + u q^2,0) \right),
\end{eqnarray}
$T_{\perp,-}^{(nf,u)}(u,\omega)=0$, where the functions 
$t_{\perp,\parallel}(u,m)$ are defined in Eqs.~(27,28) 
of~\cite{Beneke:2001at}.

\subsection{Power-suppressed amplitudes}
\label{app:a3}

Some $1/m_b$-suppressed weak annihilation 
contributions play an important role in decays to charged 
$\rho$ mesons, because they are enhanced by the large Wilson
coefficient $C_2$. In addition, power corrections may provide the 
dominant source of isospin breaking, since, as can be seen from the 
above formulae, the transverse amplitude is independent of the charge
of the spectator quark in the leading order of the heavy quark
expansion. On the contrary, all these effects are present in the 
longitudinal amplitude already at leading power. We therefore neglect 
power corrections to the longitudinal amplitude and summarize here 
the relevant expressions for isospin-breaking power corrections to 
the transverse amplitude.

{\it Weak annihilation.} 
Denoting the power-suppressed contributions to ${\cal T}_\perp^{(i)}$ 
defined in (\ref{calT}) by $\Delta {\cal T}_\perp^{(i)}$, we find 
for the annihilation terms at order $\alpha_s^0$ ($\hat s = q^2/M_B^2$)
\begin{eqnarray}
  \Delta {\cal T}_{\perp}^{(t)} \Big|_{\rm ann} &=&
  - e_q \, \frac{4 \pi^2}{3} \, \frac{f_B f_\perp}{m_b M_B} \,
   \left(C_3 + \frac43 ( C_4 + 3 C_5 + 4 C_6 )\right) 
   \, \int_0^1 du \, \frac{\phi_\perp(u)}{\bar u + u \hat s} 
\nonumber\\
&& + \,e_q \, \frac{2 \pi^2}{3} \, 
     \frac{f_B f_\parallel}{m_b M_B} \, 
     \frac{m_V}{(1-\hat s) \, \lambda_{B,+}(q^2)} \,
     C_q^{34},
\nonumber\\[0.2cm] 
   \Delta {\cal T}_{\perp}^{(u)} \Big|_{\rm ann} &=&
- e_q \, \frac{2 \pi^2}{3} \, 
     \frac{f_B f_\parallel}{m_b M_B} \, 
     \frac{m_V}{(1-\hat s) \, \lambda_{B,+}(q^2)} \,
     C_q^{12},
\label{calTuAnnih}
\end{eqnarray}
which generalizes the corresponding results  
in~\cite{Kagan:2001zk,Feldmann:2002iw}. The inverse moments 
of the $B$ meson distribution amplitudes, $\lambda_{B,\pm}(q^2)$, 
are defined in Eq.~(49,50) of~\cite{Beneke:2001at}. 

{\it Hard spectator scattering.}
The power-suppressed hard scattering terms at order $\alpha_s$ 
read~\cite{Kagan:2001zk,Feldmann:2002iw}
\begin{eqnarray}
  \Delta {\cal T}_{\perp}^{(t)} \Big|_{\rm hsa} &=&
   e_q \, \frac{\alpha_s C_F}{4\pi} \,\frac{\pi^2 f_B}{N_c m_b M_B} 
   \Bigg\{12 C_8^{\rm eff} \, \frac{m_b}{M_B} \, f_\perp
     X_\perp(q^2/M_B^2)
\nonumber\\
&& + \,8 f_\perp \int_0^1 du \, \frac{\phi_\perp(u)}{\bar u+ u \hat s} 
     \, F_V^{(t)}(\bar u M_B^2 + u q^2)
\nonumber\\
&& - \,\frac{4 m_V f_\parallel}{(1-\hat s) \, \lambda_{B,+}(q^2)} 
     \int_0^1 du \int_0^u dv \, \frac{\phi_\parallel(v)}{\bar v} 
     \, F_V^{(t)}(\bar u M_B^2 + u q^2) \Bigg\}.
\end{eqnarray}
The quark-loop function $F_V^{(t)}(s)$ is denoted 
$F_V(s)$ in~\cite{Feldmann:2002iw}, where also the 
integral $X_\perp(\hat s)$ can be found. For $q^2=0$ 
this integral suffers from a logarithmic endpoint singularity as 
$u\to 1$. We treat this singularity, which signals a breakdown of 
factorization for the power corrections, with the same ad hoc 
cutoff as in~\cite{Feldmann:2002iw}.  The corresponding expression 
for $\Delta {\cal T}_{\perp}^{(u)} \Big|_{\rm hsa}$ is obtained 
by the replacements $C_8^{\rm eff}\to 0$ and $F_V^{(t)}(s)\to 
F_V^{(u)}(s)$, where 
\begin{equation}
  F_V^{(u)}(s) = \frac34\left(C_2 - \frac16 C_1\right) \,
   \big[ h(s,m_c) - h(s,0) \big].
\end{equation}

The numerically largest power correction is 
$ \Delta {\cal T}_{\perp}^{(u)} \Big|_{\rm ann}$, because it comes 
with a large combination of Wilson coefficients 
$C_q^{12} \approx 3$ (when $q=u$, i.e.~for $B^\pm$ decay). 
In this weak annihilation effect 
the photon is emitted from the spectator quark. Therefore the matrix 
element of ${\cal O}_2^u$ factorizes at leading order in $\alpha_s$
into 
\begin{equation}
\langle \gamma^*(q,\mu) \rho^-(p',\varepsilon^*)|{\cal O}_2^u 
| B^-(p)\rangle 
= -i f_\rho m_\rho \varepsilon^*_\nu \,\langle \gamma^*(q,\mu)|
 \bar u \gamma^\nu (1-\gamma_5) b| B^-(p)\rangle
\end{equation} 
assuming the decay is $B^-\to\rho^- \ell^+\ell^-$, for which the 
effect is most important. The $B\to\gamma^*$ transition matrix element 
is dominated by hard scattering in the heavy quark limit, but 
power corrections (to weak annihilation, which is itself a power
correction to the process of interest) may be significant, especially for 
$q^2\simeq 0$~\cite{Khodjamirian:1995uc,Ali:1995uy}, where the 
hadronic structure of the photon may be resolved. To keep the 
discussion general, we parameterize the matrix element by
\begin{eqnarray}
\langle \gamma^*(q,\mu)|
 \bar u \gamma^\nu (1-\gamma_5) b| B^-(p)\rangle &=& 
i g_{\rm em} e_q \,\frac{M_B}{2} 
\bigg\{\xi_\perp^{(B\gamma^*)}(q^2)\left[i \,\epsilon^{\mu\nu\rho\sigma}
n_{+\rho} n_{-\sigma}
+2 g_\perp^{\mu\nu}\right] 
\nonumber\\
&&\hspace*{-4cm} - \, 2 \,\xi_\parallel^{(B\gamma^*)}(q^2) 
\,n_+^\mu n_+^\nu\bigg\}
+ \mbox{corrections},
\end{eqnarray}
where the vectors $n_\pm$ are given by $p^\prime =E n_-$ ($E$ the 
energy of the $\rho$ meson) and $p=M_B (n_-+n_+)/2$. The form 
factors defined by this parameterization describe the transition 
of a $B$ meson into a virtual photon of mass $\sqrt{q^2}$ 
at zero momentum transfer $p^{\prime \,2}=(p-q)^2=0$. In the heavy 
quark limit we find
\begin{equation}
\xi_\perp^{(B\gamma^*)}(q^2) = \frac{f_B}{2 \lambda_{B+}(q^2)},
\qquad 
\xi_\parallel^{(B\gamma^*)}(q^2) = \frac{f_B}{2 \lambda_{B-}(q^2)}.
\label{leading_bgamma}
\end{equation}
 The ``corrections'' stand for additional Lorentz structures that vanish 
in the heavy quark limit, and also when the leading corrections 
from the hadronic structure of the photon are included. We may
therefore adopt a more conservative treatment of the leading weak 
annihilation effects by expressing (\ref{annlong}) (and, accordingly, 
$T_{\parallel,-}^{(0,u)}$) in terms of 
$\xi_\parallel^{(B\gamma^*)}(q^2)$ and (\ref{calTuAnnih}) 
in terms of $\xi_\perp^{(B\gamma^*)}(q^2)$. The transverse form 
factor at $q^2=0$ has been estimated with 
QCD sum rules~\cite{Khodjamirian:1995uc,Ali:1995uy}, 
leading to a result not too different from the 
heavy quark limit (\ref{leading_bgamma}). In our numerical analysis 
we use (\ref{leading_bgamma}) to compute the central value of these 
form factors, but we assign a 50\%  theoretical error 
to this estimate. 

\end{appendix}

\end{document}